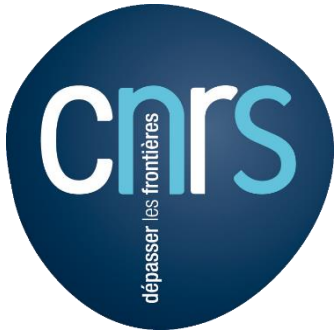 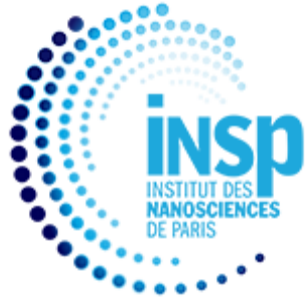 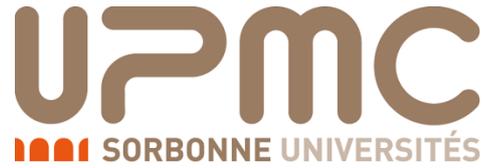

Faculté de Physique de l'Université Pierre et Marie Curie

DOSSIER POUR UNE CANDIDATURE A L'HABILITATION A DIRIGER DES RECHERCHES

Defended june 6th 2017

# Optoelectronics of confined semiconductors : the case of colloidal nanocrystals and their application to photodetection

Emmanuel Lhuillier

CR2 CNRS

COMMITEE

- **Peter. Reiss**          CEA - Grenoble, referee
- **AloyseDegiron**         C2N - Paris 11, referee
- **Christiophe Delerue**   IEMN - Lille, referee
- **Abhay Shukla**          IMPMC - Paris 6
- **Emmanuelle Lacaze**     INSP - Paris 6
- **Alexandru Nedelcu**     Sofradir - Palaiseau

E. Lhuillier's Manuscript for HDR defense – Optoelectronics of nanocrystals    2# Table of Content

Usual abreviations .................................................................................................................... 4
ACNOWLEDGEMENTS ........................................................................................................... 5
0. INTRODUCTION .............................................................................................................. 8
    0.1. CV ............................................................................................................................. 8
    0.2. PhD Work ................................................................................................................ 10
    0.3. Student supervision ................................................................................................ 13
    0.4. Main Collaborations ................................................................................................ 14
    0.5. Research strategies ................................................................................................ 15
    0.6. Manuscript organization ......................................................................................... 16
1. COLLOIDAL NANOCRYSTALS : From material to devices ......................................... 17
    1.1. Semiconductor colloidal nanoparticles .................................................................. 17
        1.1.1. Quantum confinement ..................................................................................... 17
        1.1.2. Colloidal synthesis ........................................................................................... 18
        1.1.3. Ligands ............................................................................................................. 19
    1.2. Transport ................................................................................................................. 20
        1.2.1. Hopping transport ............................................................................................ 20
        1.2.2. Transistors ........................................................................................................ 21
    1.3. Optoelectronic of nanocrystals .............................................................................. 22
        1.3.1. The nanocrystals as phosphor for display ...................................................... 22
        1.3.2. Photodetection ................................................................................................. 23
2. INFRARED NANOCRYSTALS ....................................................................................... 24
    2.1. Electronic structure ................................................................................................. 24
    2.2. Synthesis and Doping of infrared nanocrystals ..................................................... 25
        2.2.1. HgTe as interband material .............................................................................. 25
        2.2.2. HgSe as Intraband Material ............................................................................. 26
    2.3. Electronic transport in narrow band gap nanocrystals .......................................... 28
        2.3.1. Tuning the surface chemistry to boost the carrier mobility ........................... 29
        2.3.2. Dynamic aspect of transport ........................................................................... 31
    2.4. Toward a new generation of infrared detector ...................................................... 32
        2.4.1. State of the art ................................................................................................. 32
        2.4.2. Current limitations ........................................................................................... 33
3. 2D NANOCRYSTALS ..................................................................................................... 34
    3.1. Cadmium chalcogenide nanoplatelets .................................................................. 34
        3.1.1. Materials ........................................................................................................... 34
        3.1.2. An original approach for cleaning and film deposition: electrophoresis ...... 34
        3.1.3. Impact of the dimentionality ........................................................................... 36
    3.2. Transport in nanoplatelet arrays ............................................................................ 37

# Usual abreviations

IR : Infrared

QCL : Quantum Cascade Laser

QCD : Quantum Cascade Detector

CQD : Colloidal Quantum Dots

DDT: Dodecanthiol

EDT: Ethandithiol

EtOH: Ethanol

FET: Field Effect Transistor

PL: Photoluminescence

FWHM: Full Width at Half Maximum

## ACNOWLEDGEMENTS

Bien que le manuscrit soit en anglais par la suite, j'ai choisi de faire une petite entorse pour cette section. Finalement, ce manuscrit est l'occasion de faire le point sur mes 10 années de recherche et sur les multiples rencontres que ce début de carrière a pu engendrer.

Je voudrais commencer par remercier mes rapporteurs et membres du jury pour avoir pris le temps de lire et analyser ce manuscrit tout en suscitant des questions complémentaires à mon analyse.

Je dois bien sur remercier Sandrine qui semble avoir eu le nez plus fin que moi sur le choix du sujet de thèse. Peut-être par jalousie, peut-être par admiration, je suis venu marcher sur ses plates-bandes. En tout cas nous perpétuons une pratique classique dans le milieu de nanocristaux que de travailler en couple, la chimiste et le physicien.

Ce travail est bien sur indissociable de mon passage à Chicago dans le groupe de Philippe Guyot Sionnest. Il est mon mentor celui qui m'a fait éclore, me faisant passer de brave petit étudiant en thèse capable de brancher un coax entre deux appareils à un expérimentateur. Philippe m'a appris à faire. C'est-à-dire partir de pas grand-chose pour monter un instrument. Je dois avouer que mon retour en France m'a demandé une réadaptation. Je garderai en particulier en mémoire nos discussions à bâton rompu du matin. Mes remercîment américains vont également envers Sean, à qui mon expérience post doctorale doit beaucoup. Travailler avec Sean fut un vrai plaisir et notre amitié va bien au-delà du monde des nanocristaux.

Dans un style très diffèrent Benoit Dubertret a également beaucoup apporté à ce début de carrière. Premièrement, par la confiance qu'il m'a apportée, ce qui m'a permis de construire dans la durée. Auprès de lui j'ai ainsi pu apprendre que la forme est au moins aussi importante que le fond et j'ai pu développer un savoir-faire fort utile dans l'écriture de projet

Je remercie également Herve Aubin, avec qui j'ai toujours plaisir à échanger. A vrai dire ces discussions sont toujours pour moi source de beaucoup d'idées et me permettent de rester ouvert à plein de sujet dont je n'aurais même pas soupçonner l'existence. Côté ESPCI je voudrais également remercier Jérôme et Nicolas, c'est grâce à vous que j'en suis arrivé à faire de la physique du solide et du transport. Merci également à Thomas et Nicolas au moins autant pour vos connaissances chimiques que pour avoir su imposer un style si convivial au LPEM.

J'associe également à ces remerciements mes encadrants de thèse Emmanuel et Isabelle car même si les résultats obtenus sont très « niche market », j'ai pu construire et apprendre des bases solides de l'optoélectronique.

Un grand merci à mes collègues de Nexdot Solarwell. Je ne sais pas si nous aurons contribué à l'avenir de la compagnie mais en tous cas, nous avons formé une belle équipe : nous autour d'une passion commune le Mayflower. Merci à Martin pour nos débriefings du mercredi soir. Merci à Tic et Tac pour leur enthousiasme musicale à base de chanson de Carlos au laboratoire. Merci à Benmomo pour être un catalogue vivant de la synthèse de nanocristaux. Je ne peux oublier Chloé dont je dois saluer la différence de vue. Et puis il y aura eu quelques membres plus éphémères notamment les biologistes Stéphanie et Louise qui ont tenté d'accoupler streptavidin et biotin, voir les étoiles filantes comme Baptiste. Will, pour sa mythologique première semaine, doit bien sûr être intégrer à ses remerciements.

Je dois mon poste au CNRS à l'acharnement d'Emmanuelle et Bernard, qui trois années durant ont accepté de passer leur vacances de Noel à relire et corriger mon projet. J'associe à ces remerciement Sebastien Sauvage qui a lui aussi passé du temps à construire mon projet CNRS, même si l'issu ne fut pas aussi heureuse.

Je remercie également les correcteurs réguliers le Pater, et mes anglophones Jenna et Sean.

Enfin je voudrais finir ces remerciements en remerciant les différents étudiants avec qui j'ai travaillé. Par erreur, j'ai été longtemps convaincu d'aller plus vite tout seul, mais ce qui est sûr c'est qu'on va moins loin. Adrien gardera une place spéciale pour avoir été le premier. Clément et Bertille, mes deux chevaux de courses, et plus récemment Nicolas et Wasim ont leur part de culpabilité dans l'efficacité actuelle du groupe.

# 0. INTRODUCTION

## 0.1. CV

**FORMATION**

| | |
|---|---|
| 2007-2010 | **PhD from Ecole Polytechnique** for work done at ONERA, in collaboration with the laboratory MPQ (Paris VII) and Thales R&T, on electronic transport in superlattices and their application to long wavelength infrared photodetection. PhD advisor: Emmanuel Rosencher. |
| 2006-2007 | **Condensed matter physics Master II** Paris VI-Paris XI |
| 2003-2007 | **Engineering Degree from ESPCI** with double major in Physics and Chemistry |

**PROFESIONAL EXPERIENCE**

| | |
|---|---|
| 2015-present | **Research fellow (CR 2) CNRS** at the Institute for NanoSciences of Paris (INSP) at University Pierre and Marie Curie. |
| 2012-2015 | **Post doc** at LPEM – ESPCI funded by Nexdot. Main project: Optoelectronics of 2D nanocrystals. |
| 2010-2012 | **Post doc** at the University of Chicago in Philippe Guyot-Sionnest's group. Main project: Photoconduction in mid-infrared HgTe nanocrystal solids. |

**TEACHING ACTIVITIES**

| | | |
|---|---|---|
| 2015-2017 | ESPCI | Electron Microscopy Lab class (40h/year), level bac +4 |
| 2008-2010 | U. Diderot Paris 7 | « Moniteur » (2x64 h). Electromagnetism Lab class (level bac +2) and IT lab class (level bac +1). |

**PRIZES AND AWARDS**

| | |
|---|---|
| 2015 | Langlois Foundation prize |
| 2011 | Ecole Polytechnique Prize for best PhD |

**MAIN FUNDED PROJECTS**

| | |
|---|---|
| 2017-2018 | Industrial project with Nexdot for the development of an infrared photodetector based on CQD (180k€) |
| 2016-2019 | Labex Matisse - PhD thesis funding (130k€) |
| 2017-2018 | C-nano project dopQD: tuning the doping of self-doped nanocrystals (20k€) |
| 2014-2015 | Concours Mondial d'innovation: Electrochemical charging of 2D colloidal material (193k€) |

**ORGANIZATION OF SCIENTIFIC EVENTS**

| | |
|---|---|
| 2016 | Organization of a mini colloquium on colloidal semiconductor nanocrystals during the condensed matter day (JMC) at Bordeaux – 50 participants |

2011        Nanotalk: a student symposium on nanomaterial at the University of Chicago - 30 participants

## SCIENTIFIC PUBLICATIONS

- H factor 16 from Google Scholar (12 according to ISI)
- 40 publications in peer-reviewed journals (22 as first or last authors), see references: 1-40 including 5 reviews (ref 15,22,30,32,39).
- 11 conference proceedings, see references: 41-51.
- Member of the editorial board of Scientific Report.
- Regular referee for Nature Communication, Scientific Report, Advanced Material, Nanoscale, Applied Physics Letter, ACS Photonics, J Mater Chem C, Journal of Applied Physics, IEEE Journal of Quantum Electronics…
- 6 patents and patent applications, see references 52-57.
- 7 conference invited talks: 58-64 and seminar in laboratory, see references 65-76

## 0.2. PhD Work

Publications relative to this work: 1-7,11.

Conference proceedings relative to this work : 41-43.

I conducted my PhD work at ONERA under the supervision of Emmanuel Rosencher. This project was also undertaken in collaboration with laboratory MPQ (Paris VII) for the modelling side. Thales R&T was our sample provider. This work was dedicated to the study of electronic transport in semiconductor superlattices and its application to long infrared wavelength detection. The research included both experimental and modeling aspects. The basic idea was to measure and model the electronic transport properties of the GaAs/AlGaAs heterostructure, operated in the tunnel regime, and to determine what was the transport bottleneck in order to suggest improved semiconductor designs.

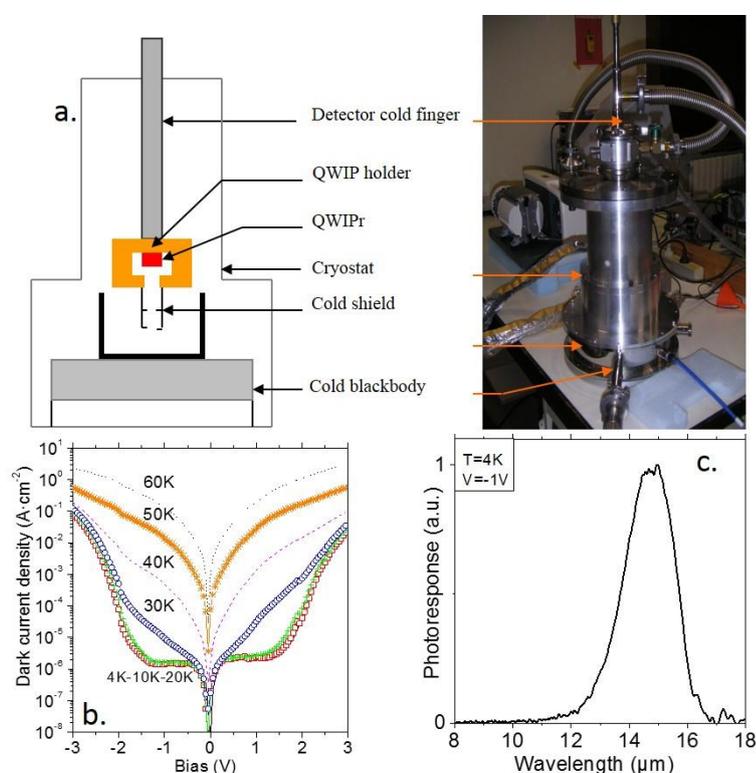

On the experimental side, I measured, in cryogenic condition, the dark current of a multiquantum well heterostructure (QWIP) dedicated to 15µm photodetection (see Figure 1c). 15 µm infrared emission corresponds to the maximum emission of a black body at ≈200 K. This device is typically made of 40 doped quantum wells of GaAs surrounded by a barrier of AlGaAs, with a 15% content of Al. To obtain narrow energy transition, fairly large (7.3 nm) wells have been grown by molecular beam epitaxy (MBE). Each well includes two steady states—one ground state ≈40 meV above the bottom of the well and one almost resonant with the barrier. Under IR light, electrons are promoted to the excited state and can then easily drift in the continuum. Such a long wavelength device is dedicated to infrared detection under low photon background.

Figure 1 a. Scheme and image of a dual cryostats setup dedicated to the characterization of long wavelength QWIP operated under low background photon flux. b. Dark current density as a function of applied bias for the 15µm QWIP investigated in ref 4. c. Spectral photoresponse of the 15µm QWIP investigated in ref 4.

In addition to photocurrent, a dark current is also flowing in the semiconductor stack. At high temperatures, it comes from the phonon activation of the carrier up to the continuum. At low temperatures, tunnel transport through the barrier also occurs, and determines the ultimate performance of a given heterostructure[1].

Experimentally, I developed a setup dedicated to the measurement of dark current and the photoresponse of the QWIP. Measuring dark current can be quite challenging, since it relates to low level of current (sub pA) (see Figure 1b). To measure the photoresponse, I built a two-cryostat setup;

one is used to cool down the QWIP to 4 K, and the other one is used as cold blackbody (77 K-300 K) (see Figure 1a). This setup was used to measure a calibrated photoresponse of the QWIP over 4 orders of magnitude of incident photon flux.

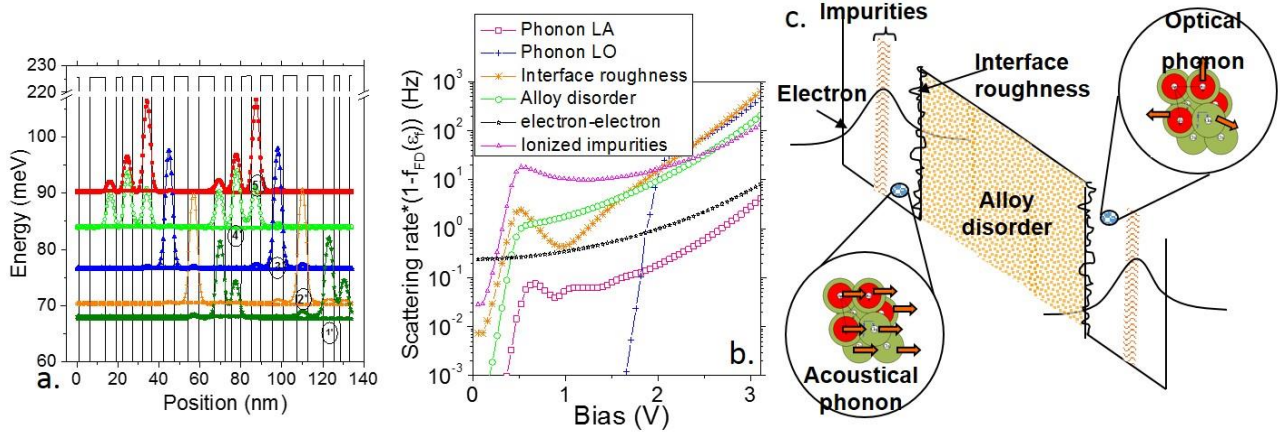

*Figure 2 a. Band structure and wavefunction of a THz quantum cascade detector, as proposed in ref 3. b. Scattering rates resulting from different scattering processes as a function of the applied bias for the 15µm QWIP structure, as investigated in ref 4. c. Different scattering mechanisms considered in the modelling tools developed in ref 3-5, 7.*

The obtained results were then used as experimental input and compared with the results obtained from my modeling tool[2,4]. I developed from scratch a simulation code dedicated to the determination of electronic structure and transport properties of semiconductor heterostructures. In the first step, a 1D heterostructure profile (energy and mass) is defined (see Figure 2a). Then, using a kp code, the time independent 1D Schrodinger equation $\left[-\frac{\hbar^2}{2}\frac{d}{dz}\frac{1}{m^*(z)}\frac{d}{dz}+V(z)-eFz\right]\xi(z)=E\xi(z)$ is solved and the steady states of the heterostructure are determined. This gives us access to the electronic spectrum and wavefunction of the system. I then use the Fermi golden rule $\Gamma(K_i)=\tau^{-1}_{tunnel}(E,F)=\frac{2\pi}{\hbar}\sum_{Kf}|\langle f|H|i\rangle|^2\delta(\varepsilon_i-\varepsilon_f)$ to evaluate the scattering rate between two states coupled by different Hamiltonians. These interactions are the electron–phonon interaction (acoustic + optical), the coulombic interaction between electrons and ionized impurities (i.e., doping), and the scattering resulting from interface roughness and alloy disorder (see Figure 2c). This type of "exact" modelling is only made possible by the almost perfect quality of the GaAs/AlGaAs heterostructure, which makes transport limited by fundamental processes. Finally, the current is evaluated through the expression $J=\int_{E_1}^{\infty}e\frac{m^*}{\pi\hbar^2}.\tau^{-1}(E,F).(1-f_{FD}(\varepsilon_f)).f_{FD}(\varepsilon_i)d\varepsilon_i$ which includes the thermally activated population of each state. Such modeling of the current is actually very common for a quantum cascade laser,[77-79] but its use for QWIP was previously lacking. The actual challenge comes from the fact that quasi-resonant states already exist in the detector, and so a steady-state evaluation of the wavefunction tends to overestimate the state coherence. I thus proposed a model which introduces a coherence length to overcome this difficulty[4].

This modeling tool was used to determine which interaction is leading to the main current contribution. In the 15 µm QWIP, the bottleneck results from the interaction of electrons with ionized impurities. This suggests that the current doping profile located in the center of the well, which has a strong overlap with the electron wave function, is actually maximizing the scattering rate relative to the election doping interaction. I thus suggested a new doping profile, shifted from the well center. The Sample have been grown by Thales. We measured a 15 % reduction of the dark current[7], which is not so small given the high maturity of this QWIP structure.

I also used my simulation code to evaluate the impact of growth defects on the final optoelectronic properties of QWIP[5]. I considered their effect on non-sharp interfaces and inhomogeneous dopant distribution and investigated their impact on the electronic spectrum. Another interesting output of the simulation tool was its use in understanding the transport properties of THz quantum cascade detectors[3,80] (QCD). QCDs are the detecting counterpart of the quantum cascade laser[81]. They have been initially proposed in the mid-infrared,[82,83] and in this range of energy the coupling between the states of the cascade occurs through optical phonon. However, this type of coupling is no longer possible in the THz since the whole states ladder is included in a small range of energy below the optical phonon energy (36 meV in GaAs). Thus the coupling has to occur through a different scattering process. The interface roughness was determined to be the main contribution[3]. The estimation of the two driving parameters relative to this interaction was critical. The interface length and the coherence length between defects was then determined from TEM imaging conducted in a dark field condition by Gilles Patriarche at LPN.

Overall this PhD work was great training, which gave me knowledge on both the experimental and modelling side. The great maturity of the field on the modelling and material side ensure that properties of devices can be evaluated with a high level of accuracy and a limited number of tunable parameters. However, this high maturity, combined with the lack of tunability of the sample, can be frustrating. Consequently, when I chose my post doc experience it was important for me to get greater control over the sample fabrication. Colloidal quantum dots (CQD) have perfectly answered this demand.

### 0.3. Student supervision

I have supervised 8 students at levels below PhD (Master II, technician, undergraduate) and one post doc. I co-supervised one finished PhD thesis and currently co-supervise two of them.

- **PhD Student**

| Student | Level /place | period | Thesis title - co supervisor |
|---|---|---|---|
| Adrien Robin | ESPCI | 2013-2016 Defended 4/11/16 | Opto-électronique de boîtes et puits quantiques colloïdaux – Application au photo-transport<br>Benoit Dubertret |
| Clement Livache | INSP/ESPCI | 2016-2019 | Dynamic aspect in narrow band gap nanocrystals<br>Benoit Dubertret |
| Bertille Martinez | INSP | 2016-2019 | Control of carrier density in narrow band gap nanocrystals<br>Emmanuelle Lacaze |
| Wasim Mir | INSP | 2017 | Visiting student from Angshuman Nag's group |

- **Post doctoral student**

| Student | Level /place | period | Project |
|---|---|---|---|
| Remi Castaing | ESPCI | 2015 – 6 months | Electrochemistry of 2D colloidal nanocrystals |
| Nicolas Goubet | INSP | 2017-2018 | Mid IR nanocrystal synthesis |

- **Pre PhD Students**

| Student | Level /place | period | Thesis title - co supervisor |
|---|---|---|---|
| Paul Rekemeyer | Undergraduate /U. Chicago | 2011 6 months | Noise in nanocrystal solid<br>Philippe Guyot Sionnest |
| Adrien Robin | Master II/ ESPCI | 2013 5 months | Transport in nanoplatelet arrays<br>Benoit Dubertret |
| Daniel Thomas | Master II/ ESPCI | 2014 5 months | Transport in nanotrench devices<br>Benoit Dubertret |
| Marion Scarafagio | Master II/ ESPCI | 2015 5 months | Infrared nanocrystal<br>Benoit Dubertret |
| Patrick Hease | technician | 2013-2014 | Electrophoresis of nanoplatelets<br>Benoit Dubertret |
| Loic Guillemot | Master II/ ESPCI | 2015 5 months | New material obtained from self-assembly of colloidal nanocrystals<br>Emmanuelle Lacaze |
| Sharif Shahini | Master II/ INSP | 2011 2 months | Ferro electric gating of nanocrystal film<br>Emmanuelle Lacaze |
| Clement Livache | Master II/ INSP | 2011 6 months | Transport in narrow band gap nanocrystals<br>Benoit Dubertret |

For seven of them their work has led to a publication.

## 0.4. Main Collaborations

| Collaborator | Intitutions/compagnies | Topic |
| --- | --- | --- |
| S. Ithurria | LPEM – ESPCI | 2D Nanocrystal synthesis |
| H. Aubin | LPEM - ESPCI | Tunnel transport |
| A. Ouerghi | LPN -CNRS | 2D system : graphene and TMDC |
| J.-F. Dayen | IPCMS – U. Strasbourg | Nanofabrication and electronic transport |
| B. Dubertret | Nexdot | Nanocrystal synthesis and infrared |
| L. Biadala | IEMN (U. Lille) | Sample provider |
| J. Houel | ILM (U. Lyon) | Sample provider |
| V. Krachmalnicoff | Institut Langevin | Sample provider |
| Q. Glorieux | LKB | Sample provider |

## 0.5. Research strategies

My research has always been at the interface between fundamental science and more applied research. The early motivation is generally driven by applied needs, and in particular in the field of infrared photodetection, by the need to understand current performance limitations and to identify strategies for performance enhancements. To achieve this goal, it is necessary to go fairly deep into the understanding of the material and its electronic structure. That is also why my research is always multidisciplinary, encompassing a material understanding of the electronic structure in question, the measurement of its transport properties, all the way up to the performance of the optoelectronic device.

My interest in nanocrystal synthesis was initially driven by the need to obtain samples. But in this field in which materials are still not mature, the control of chemistry is always necessary to tune physical parameters (e.g., doping, height and length of a tunnel barrier). This might not be the case a decade from now, because synthesized CQD have an increasing complexity, but as of yet it is still possible to do research from the CQD synthesis up to the application. This gives us tremendous advantages in terms of material control.

Another key difference from the III-V semiconductor that I studied during my PhD is the very limited understanding of the electronic structure. In III-V semiconductors, all parameters are known within several digits of accuracy. This is not the case for II-VI semiconductors, particularly once quantum confinement is added. Determining the physical parameters of these new materials is still a necessary challenge that has to be tackled.

I also think that CQD-based devices can also benefit from the progress of nanofabrication. The overall geometry of CQD-based devices remains simple, because the topic has been driven by groups with a strong chemical background. For my research, I will introduce strategies for the design of devices on a smaller scale, while introducing mechanisms allowing higher control of the material properties.

## 0.6. Manuscript organization

The manuscript is organized over five main parts. I first introduce colloidal nanocrystals, their synthesis, the transport properties in nanocrystal solids, and their applications. The second part is dedicated to narrow band gap nanocrystals and their application to photoconduction and detection in the mid-infrared. In particular, the document includes my contribution to the use of HgTe and HgSe CQD as mid-IR photoconductors. This research was initiated by the time I was a post doc in the Guyot-Sionnest group, with Sean Keuleyan (graduate student) and Paul Rekemeyer (undergraduate student) as co-workers. The contribution concerning HgSe is more recent, and was supported by two former students (Patrick Hease and Marion Scarafagio) and two ongoing PhD student (Bertille Martinez and Clément Livache).

Part three deals with transport and phototransport in 2D nanoplatelet arrays. This research corresponds with my period in residence at LPEM and overlaps with the PhD project of Adrien Robin. Daniel Thomas, a Masters student, was also involved for the nanoscale device part.

The fourth part is focused on Van der Waals heterostructures with mixed dimensionality. This work was initially driven by the idea to decouple transport and absorption in order to overcome hopping transport. To do so, we coupled nanocrystals with a graphene layer to take advantage of its large carrier mobility while the absorption is occurring in the semiconductor nanocrystals. This research provided the opportunity to initiate a collaboration with Abdelkarim Ouerghi at C2N. Since then, we have amplified our collaboration with several projects dedicated to transport in Van der Waals heterostructures.

In section five, I describe the main perspective of my work for the upcoming years. This includes the ongoing PhD projects of Bertille Martinez and Clément Livache and post doc by Nicolas Goubet. I suggest that I will pursue research in three main directions, including the synthesis of new narrow band gap materials, the probing of their electronic structure, and the investigation of their relaxation dynamic. Finally, I propose the design of a new generation of optoelectronic devices based on CQD as active material.

# 1. COLLOIDAL NANOCRYSTALS : From material to devices

For a review, in French, on the topic see ref 32.

Up to the 1960's, low dimensional objects were only limited to theory. Great progress in material science has made confined material a reality. The progress of epitaxy has led to the development of quantum well[84] (2D) and then quantum dot[85] (0D) structures. Additionally, progress in micro- and nanofabrication also made possible the definition of confined areas thanks to electrostatic gating on 2D electron gas. These new materials are not limited to the academic world and some of them have reached a mass market. This is in particular the case of the GaAs quantum well, which is now a key building block for lasers and the quantum cascade laser[77] is probably the most achieved example of quantum engineering. Achieving confined objects is nevertheless not limited to these approaches, and chemical paths were developed by the end of the 1980's. At the early beginning of the field, barely nm scale particles were synthetized with a broad size dispersion. A key breakthrough occurred in 1993 in the Bawendi group, where the hot injection method was developed.[86] It has led to monodispersed nanoparticles with an atomic-like spectrum. For these nanoparticles, the inhomogeneous broadening is low enough to observe a structured spectrum resulting from the sparse density of state of 0D semiconductor nanoparticles. Over the following 25 years, these so-called quantum dots generated a huge research effort, not only on the material side but also with respect to their applications.

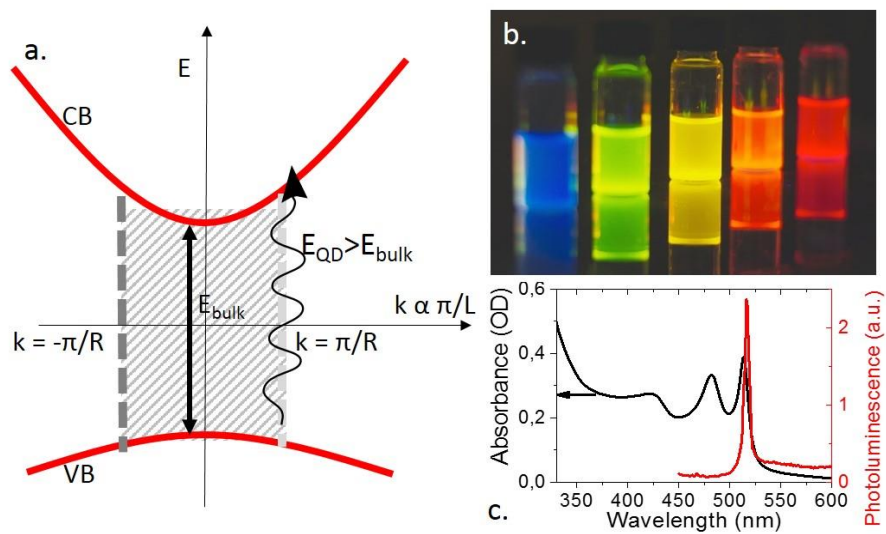

Figure 3 a. Band structure of a confined semiconductor. b. Image of a nanocrystal solution of cadmium chalcogenides with different levels of confinement. c. Absorption and emission spectrum of CdSe nanoplatelets.

## 1.1. Semiconductor colloidal nanoparticles
### 1.1.1. Quantum confinement

Colloidal nanocrystals, also called quantum dots, are nm size semiconductor nanoparticles. Their size is typically below the Bohr radius. The latter is given by $a = \frac{\varepsilon_0 \varepsilon_r h^2}{\pi \mu e^2}$ with $\varepsilon_0$ the vacuum permittivity, $\varepsilon_r$ the material dielectric constant, $h$ the Planck's constant, $\mu$ the reduced effective mass and $e$ the proton charge. The Bohr radius represents the mean spacing between a photogenerated electron hole pair. Once the size of a nanoparticle is reduced below this size, quantum confinement appears.

Only sizes between that of the atomic lattice up to that of the nanocrystal can now be reached. In the reciprocal space this corresponds to momentum from the infinite down to $\pi/R$ with $R$ as the CQD radius (see Figure 3a). The band edge energy is thus corrected toward higher energy and it results in a blue-shift of the optical features. Tuning the size of the cadmium chalcogenides gives access to the whole visible spectrum (see Figure 3b).

The sparse density of state of CQD results in a structured optical spectrum. Typically, the absorption spectrum presents different peaks close to the band edge energy. They result from different possible transitions (see Figure 3c). At higher energy, the absorption becomes far less structured and is related to the amount of absorbing semiconductor, independent of the nanoparticle size and shape. The photoluminescence (PL) of these CQDs is the source of the interest in this material. The PL occurs at an energy slightly lower than the absorption band-edge. The difference between the absorption and the emission peak is called the Stoke-shift, which depends on the CQD polydispersity and possible trap states within the band gap.

### 1.1.2. Colloidal synthesis

CQDs are synthetized in solution. A typical setup is shown in Figure 4a. During a typical synthesis, cation salts are introduced with ligand and solvent. The solution is degassed to get rid of dissolved water and oxygen. To increase the temperature, the atmosphere is switched to argon or nitrogen. The temperature is set to a value which strongly depends on the targeted material: 250°C for CdX, 150°C for PbX, and 80°C for HgX. Here X is a random chalcogenide (S, Se and Te). By the end of the synthesis, an excess of ligand is introduced to passivate the CQD surface.

The obtained nanoparticles are crystalline, as shown on the high resolution TEM image, which highlights the lattice fringes (see Figure 4b). In addition, this nanoparticle is capped by an organic shell made of ligand molecule (see Figure 4c). With now more than two decades of research in the field, a broad range of material and shapes can be obtained in nanocrystal form (see Figure 5). This includes metals (Au), wide band-gap (ZnO, CdSe, CsPbBr$_3$) and narrow band gap (PbS, PbSe, Sb$_2$Te$_3$), semiconductors, semimetal (HgTe, HgSe), superconductors (Pb).

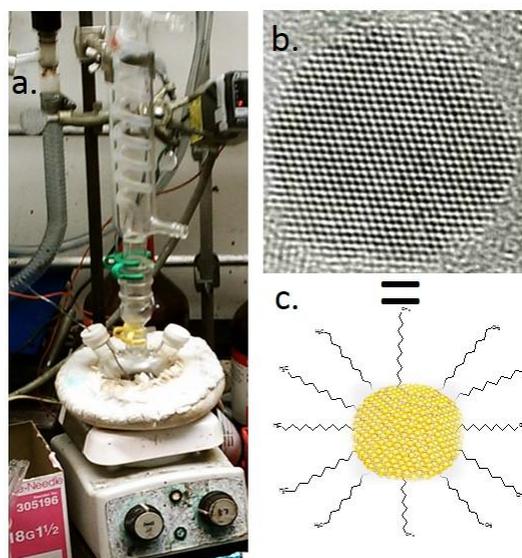

*Figure 4a. Image of a colloidal synthetic setup, from PGS' lab. b. TEM image of a single nanocrystal. c. Scheme of a nanocrystal with its inorganic core and its surrounding ligand shell.*

Control of the shape and dimensionality has reached a high level of maturity, enabling production of spheres, rods, wires, platelets, cubes and stars. In what follows, my research deals with the appropriate choice of material from this library in order to build devices with enhanced performance, or, alternatively, in order to probe material properties.

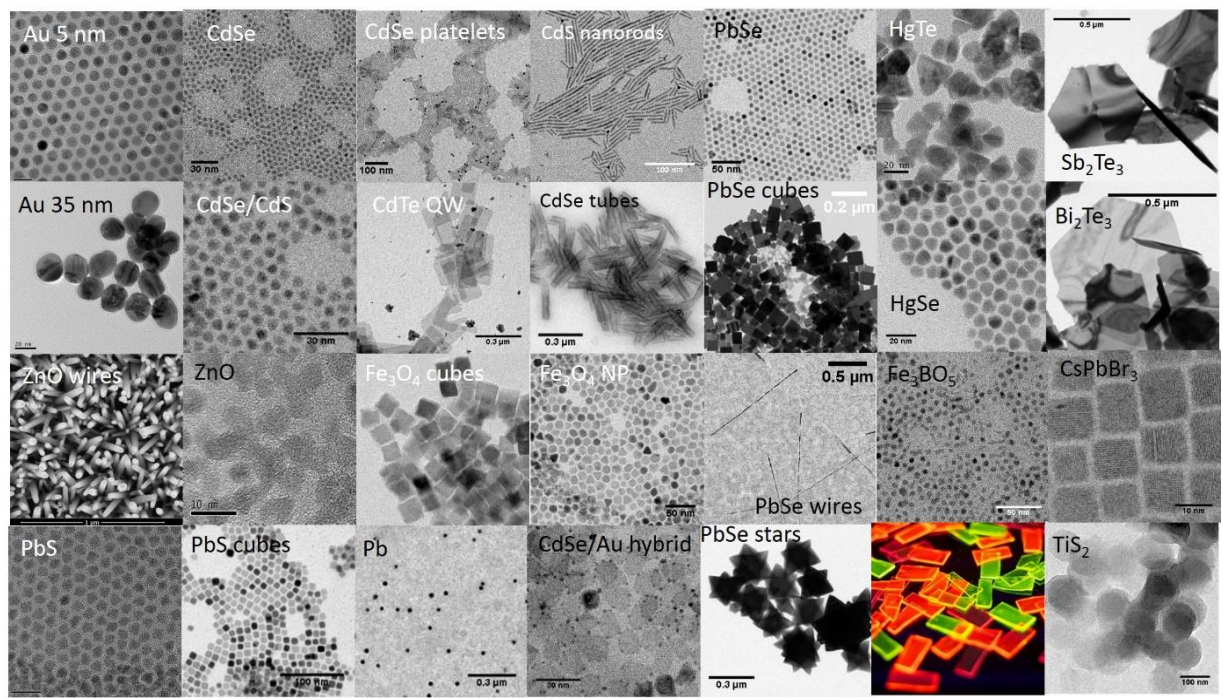

*Figure 5 TEM image of nanocrystals of various composition and shape that I have synthesized.*

### 1.1.3. Ligands

Ligands are critical elements of nanocrystals. They typically are made of an organic functional group such as carboxylic acid, amine, thiol or phosphine, which bond to the nanocrystal surface and to an alkane chain. The length of the chain typically ranges between 12 to 18 carbons.

Ligands have three roles. (i) During the synthesis they make the nanocrystal surface difficult to access for additional precursors and thus slow down the reaction rate. This is critical to preserve the nm size of the quantum dots. Their introduction is the main difference from previous synthesis in the 1980's, which was conducted in aqueous media. (ii) Ligands also preserve the colloidal stability of the nanoparticles. The non-polar nature of the alkane chain favors nanocrystal solubilization in non-polar solvents. (iii) Finally, ligands play a key role in the electronic passivation of the surface states. Dangling bonds tend to introduce states with the semiconductor band gap. The hybridization of these surface states with the ligands is used to push these states away from the gaps.

In other words, their presence is mandatory but they actually are very detrimental to most nanocrystal applications. For biological applications, water soluble CQDs are desired, but as they are at the end of the synthesis, CQD are not soluble in water. In devices, ligands behave as tunnel barriers. From STM their height has been estimated to be 2 eV, while their length corresponds to the ligand lengths (1-2 nm typically).[87] As a result, a film of nanocrystals is electrically insulating. Typical mobility is below $10^{-6}$ cm$^2$V$^{-1}$s$^{-1}$. Ligand exchange procedures have thus been developed to reduce ligand length. Initially the procedure was developed directly on the film. A pre-formed nanocrystal film is dipped in a solution of short ligands diluted in a non-solvent. For example, ethandithiol (EDT, a short chain with two thiol functions) is dissolved in ethanol at a 1% in volume concentration. Since ethanol is not a solvent for the CQDs, they do not get dissolved. The excess short ligands replace the initial long ligand and the related tunnel barrier shrinks from 1.5 nm to 0.5 nm, which leads to an increase of mobility in the $10^{-3}$-$10^{-2}$ cm$^2$V$^{-1}$s$^{-1}$ range. After 2009, inorganic ligands were introduced[88]. They introduced a new way to tune the inter CQD tunnel barrier while manipulating the barrier height. However, these metal chalcogenide complexes were still difficult to prepare and use. As a result, in the early 2010's, ionic short ligands were introduced. They typically consist of anionic halide[89] (Cl$^-$, Br$^-$

, I$^-$) or chalcogenides[90] (S$^{2-}$, Se$^{2-}$ and Te$^{2-}$) anions. Other anions such as SCN$^-$,[91] OH$^-$ and close derivatives of previously mentioned anions (HS$^-$) have also been proposed. These anions are both easy to use, commercially available, and allow for tunnel barrier height tunability. These new surface chemistries have made it possible to reach mobility in the 1-10 cm$^2$V$^{-1}$s$^{-1}$ range.[91,92] In this case the ligand exchange is conducted in liquid phase. Typically, some of the ions are dissolved in a polar solvent (DMF), while a top phase is made of CQDs in a non-polar solvent. After sonication the short ions cap the CQDs. They thus get transferred to the polar phase. After a cleaning step to remove the excess of ligands, a stable solution of ionically passivated CQD is obtained. These methods will be intensively used in the following sections of this report. The current mobility record[93] for CQD-based film is around 400 cm$^2$V$^{-1}$s$^{-1}$. Paradoxically the idea to use conjugated chains as ligand has so far not led to any successful results. The reason for this failure is not clear and might be due to the difference between the energy of the CQD and the polymer HOMO LUMO energy.

### 1.2. Transport
#### 1.2.1. Hopping transport

In a CQD solid transport occurs through a hopping process. Due to the difficulty involved in doping nanocrystals, charges are usually generated through optical pumping[94] or field effect[95]. Carriers then hop from CQD to CQD. At each step the carrier has to overcome a tunnel barrier.[87] The tunnel barrier, as described above, is the result of the ligand. Moreover, an imperfect surface passivation will lead to the presence of surface traps, which act as recombination centers. This results in a complex multiscale percolation path. At the nanoscale level, the CQD size distribution makes it such that large CQD are less confined and behave as traps. At the mesoscale level, the presence of cracks, which typically result from volume reduction after the ligand exchange, also makes the path to reach the electrodes even more tortuous. As we deal with transport in nanocrystal solids, all these difficulties have to be considered. I will describe some strategies I have developed to deal with these problems in the following sections.

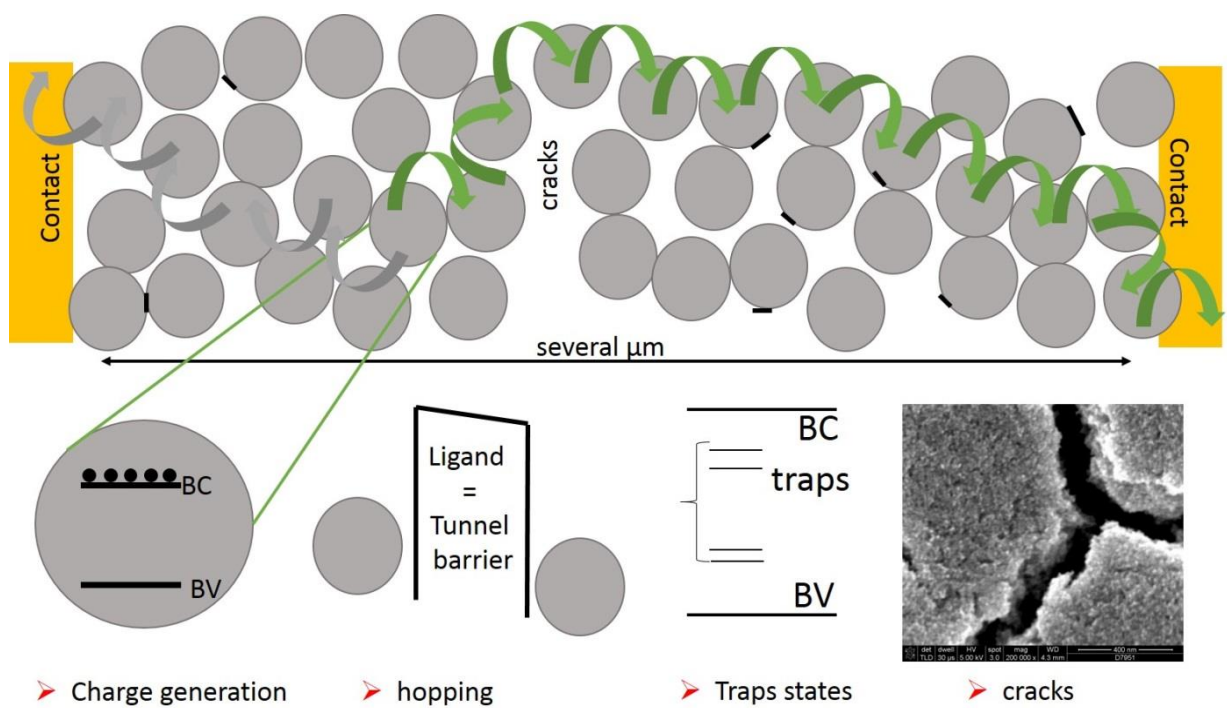

*Figure 6 Scheme illustrating the hopping transport occuring in a CQD array. Charge has to be generated within a CQD by light illumination of field effect. This charge will then hop from CQD to CQD and has to overcome a tunnel barrier at each event. Traps surface states may be introduced in*

*the gap and leads to recombination. Finally this hopping process is multi scale diffuse transport. CQD size distribution and crack leads to a complex percolation path.*

### *1.2.2. Transistors*

While CQDs have historically attracted interest for their luminescent properties, their use for transport only began in the early 2000's. This delay is a consequence of the previously mentioned need to change the CQD surface chemistry in order to make the CQD solid conductive. Research on transport in CQD solids has been motivated both by the design of devices and by its potential to be a new window onto electronic structure. Due to the inherent low mobility of CQD film, Hall geometry, which is in solid state physics the most common way to access the carrier density and mobility, is extremely difficult and only a few recent results have been obtained[96]. As a consequence, the field effect transistor (FET) has become the most popular way to conduct transport measurements in CQD solids. Clearly, CQD-based transistors are not built with the intention to replace silicon electronic. The FET is, on the other hand, the most reliable way to identify the majority carrier and the mobility of a CQD film. The latter is proportional to the derivative of the transfer curve (drain current vs gate voltage). Over the past decade, strong efforts have been made to demonstrate large mobility value. This is because mobility is the macroscopic measurement of the local inter-CQD coupling: $\left|\left\langle \psi_i | \psi_{i+1} \right\rangle\right|^2$ where $\psi_i$ is the wavefunction associated with the i[th] CQD. CQD-based FET have been initially proposed by Talapin and Murray,[95] using a $SiO_2$ dielectric layer as gate. In this case, conventional oxidized Si wafers are used and the oxide layer is used as gate insulator. This approach is far from ideal because of hysteresis and large leakage, but is very easy to develop. A ligand exchanged film of CQD is connected to drain source electrodes (see Figure 7a). As drain bias is applied, the bands are biased and the charges can drift toward the electrodes (see Figure 7b). If a gate bias is applied, surface charges are generated on each side of the dielectric. These charges can be used to move the Fermi level within the semiconductor (see Figure 7c).

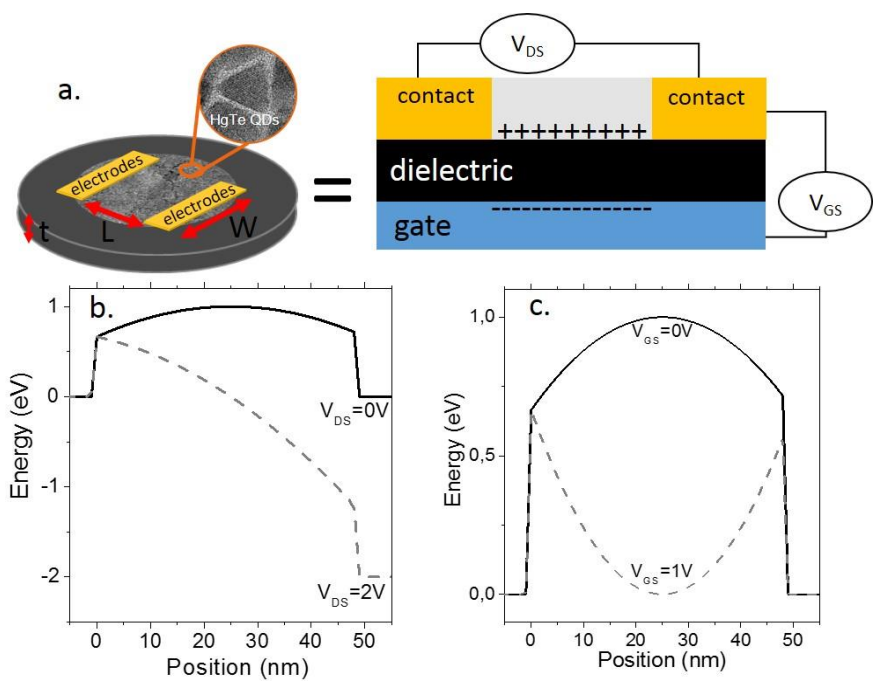

*Figure 7a. Scheme of a CQD-based transistor. 7b and c illustrate respectively the effect of drain and gate bias.*

### 1.3. Optoelectronic of nanocrystals

CQDs have generated interest in a broad range of applications which span from bio imaging[97] to optoelectronics[32,98]. I will not review all of them; I will instead focus on two applications with which I have experience.

#### 1.3.1. The nanocrystals as phosphor for display

Because of their high PL quantum yield, the use of CQD as a photon source has driven most of the research efforts. CQDs combine several advantages, such as a good resistance to photobleaching, a large absorption cross section, a broad band absorption, and a tunable PL. They can even be used as single photon source. The idea to use them as light source is therefore appealing, but the reality remains challenging. To date, the electroluminescence (EL) of CQD thin film has been demonstrated[99-101]. However, the efficiency and lifetime of such film remains technologically non-competitive. In addition, EL combines the difficulties of both luminescence and charge management. Industry has come to the conclusion that CQD-based EL is not mature yet, and have chosen to use CQD as down converters. For the last three years, CQD have been commercially available and their first mass market use is in the form of phosphor for display. In a conventional LCD LED display a blue GaN/InGaN diode emitting at 460 nm is pumping a Ce:YAG yellow phosphor (see Figure 8a). This combination of blue and yellow generates a white color. This is then filtered by a RGB filter to make colored pixels. The gamut, which is to say, the number of colors that can be reproduced by a display, strongly depends on the narrowness of the source (see Figure 8b). This is in particular the case with green, to which the eye is strongly sensitive. The narrowness of the emission is one of the key motivations for OLED. However, OLED remains expensive and its lifetime is still too short. Several industry players (Sony, Samsung) have chosen to preserve the well-established LED LCD configuration and just replace the broad Ce:YAG emitter with green and red CQD (Figure 8c). This leads to significantly improved gamut. Actually, all the challenges come from the need for good encapsulation of the CQD to make them sustain the operating conditions. The next generation of LED-pumped CQD will be operated under large photon flux (10-100 W.cm$^{-2}$, which is a tenth of the transition saturation) and high temperature (≈100°C). While I was at ESPCI, funded by the start-up Nexdot, I had the opportunity to work on this topic pursuing the idea that 2D nanoplatelets can become the next generation of phosphors and lead to an even higher gamut (see Figure 8c). This field already represents a market of 100 million to one billion dollars per year. Industry developers have massively bet on this technology and nanomaterial, and they now aim for more applications of these CQD. I am convinced that optoelectronics and, in particular, infrared photodetection, will be part of this new market.

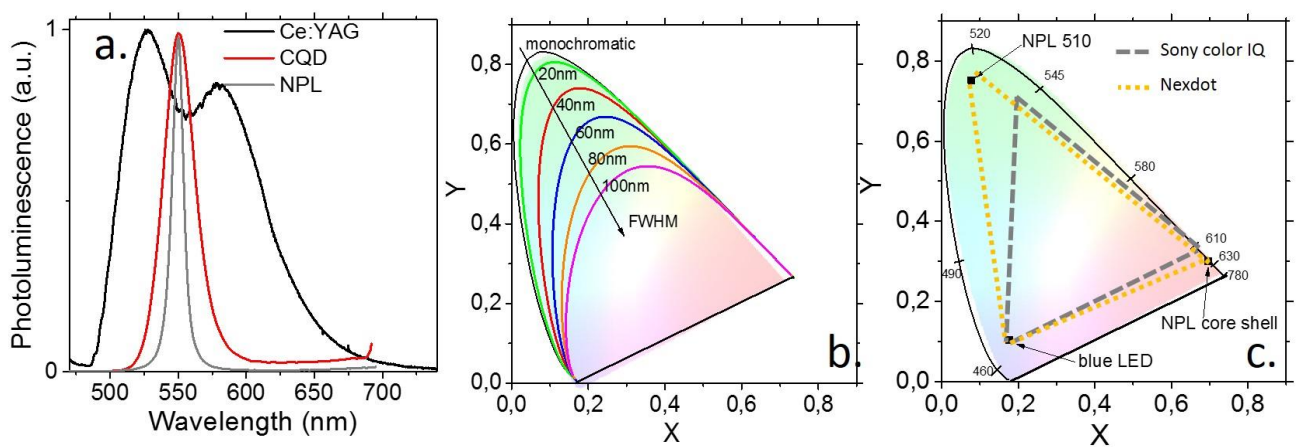

E. Lhuillier's Manuscript for HDR defense – Optoelectronics of nanocrystals    23*Figure 8a. Luminescence spectra of Ce doped YAG, CdSe CQD and CdS NPL. b. Chromaticity diagram for different FWHM of the source. c. Area of the chromaticity digram which can be accesed by the QCD color IQ technology by Sony and while using NPL as source.*

### *1.3.2. Photodetection*

It took 10 years after the introduction of nanocrystals to obtain transport in CQD solids. This progress paved the way for the integration of CQD as a building block for optoelectronics and prompted the idea to revisit all semiconductor physics using CQD as a building block. However, it was and remains a challenging question, not least because the study of transport properties is still less mature than that of bulk material. There are also more open avenues for continued research, including nanocrystal doping and developing reliable material processability. Amongst this tremendous amount of work conducted on CQD solids, the other main optoelectronic application that has driven my research on CQDs is the design of a photodetector specifically dedicated to infrared detection. This application is motivated by the fact that CQD can offer an interesting alternative to current technologies based on epitaxially-grown semiconductors for the design of low cost detectors. Indeed, CQD-based technology removed the constraint of epitaxial growth, as the material can be deposited on any substrate. We also benefit from the relative ease of processing material developed for organic electronics (spin coating, ink jet…). The challenge is to demonstrate a sufficiently high level of performance to demonstrate that CQD can indeed lead to a new generation of infrared sensors. In the following I illustrate several of my contributions to this field.

## 2. INFRARED NANOCRYSTALS

Students involved in this work: Adrien Robin, Marion Scarafagio, Patrick Hease.

Publications relative to this work: 8,9,10,12,14,26,31,35,37.

Reviews related to this work: 15,37.

Conference proceedings relative to this work: 44-47.

Patents relative to this work: 52,57.

In the visible range of wavelengths, current optoelectronic devices have succeeded in combining low cost and high performance. In the infrared this is still not the case. Infrared detection relies on two main types of detectors. Quantum (i.e., photon) detectors which are fast, but cooled. These detectors are based on narrow band-gap semiconductors such as InSb, HgCdTe or wide band gap semiconductor heterostructures (GaAs/AlGaAs or InAs/GaSb). Alternatively, energy detectors such as bolometers can detect infrared signals without cooling but at the price of a longer time response. Because of their low temperature operation, which involves cryogenic conditions, quantum detectors are dedicated to high performance needs. Their current price of $10-100 k per camera limits their use to defense and astronomy applications. There is nevertheless is a growing demand for IR detectors in applications such as building, thermal management, or night assistance car driving. These new applications require a cost disruption that current quantum technologies are unlikely to bring. I am convinced that CQDs can offer an interesting alternative to the bolometer technologies.

To build an infrared detector from a CQD film, we first need a material which is absorbant in the IR. So far in the field of CQDs most of the effort has been focused on cadmium chalcogenides because of their tunable band gap in the visible. III-V semiconductors such as InP have been more recently investigated as a low toxicity alternative, even if they remain less mature (i.e., broader FWHM, lower PL efficiency). Lead chalcogenides with their moderate band gap have also attracted a significant research effort for their use as active material for CQD-based solar cells[102]. However, the bulk band gap of these materials (0.4eV for PbS) prevents their use in the mid-infrared and thus in thermal imaging. Narrower band gap semiconductors (SnTe, $Bi_2Te_3$) and semimetals (Bi, Sn, HgTe, graphene) have to be used to push the optical absorption above 3µm. 3µm is indeed an interesting threshold. First, it corresponds to the value above which the thermal emission (i.e., blackbody emission) prevails over the reflection from warmer source (i.e., the sun). Secondly, to conduct thermal imaging, the atmosphere has to be transparent and there are two transparency windows, which are the 3-5 µm range (MWIR) and the 8-12 µm range (LWIR).

To answer this challenge HgTe was as of 2010 the most mature candidate. As a II-VI semiconductor it benefits from experience gained from CdX, and its proximity to the bulk alloy HgCdTe ensures that its band structure is reasonably well-known.

### 2.1. Electronic structure

Under bulk form, HgTe is a semimetal with an inverted band structure (see Figure 9a). The $\Gamma_6$ band which has generally the symmetry of a conduction band (in CdTe, for example) is below the $\Gamma_8$ bands which usually have the valence band symmetry. This inversion of the band is responsible for the negative value of the band gap reported for HgTe and HgSe. When they are intrinsic materials, the Fermi level lies between the two $\Gamma_8$ bands. As a result the weakly dispersive heavy-hole band with $\Gamma_8$ symmetry plays the role of the valence band, while the conduction band is the second $\Gamma_8$ band and has a similar symmetry to the light hole band in CdTe. Interband transitions occur between these two $\Gamma_8$ bands. A calculation of the electronic structure of HgTe CQD using a tight binding method (see Figure 9b) also leads to a good correlation with experimental results. [103,104]

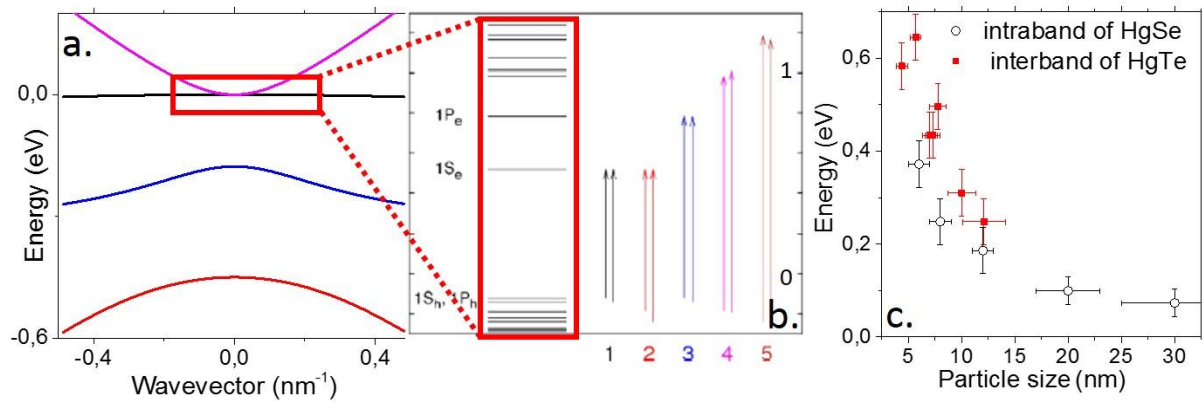

Figure 9 a. Band structure of HgSe. The two top bands have a $\Gamma_8$ symetry and the fermi level lies between these two bands. The blue lines represent the state with $\Gamma_6$ symetry. Its energy below the $\Gamma_8$ bands is characteristic of the inverted band structure of HgTe and HgSe. b. Electronic structure of a HgTe CQD determined by a tight binding approach (see ref 104). c. Size dependence of the interband energy transistion of HgTe CQD and intraband energy transistion of HgSe CQD.

## 2.2. Synthesis and Doping of infrared nanocrystals

### 2.2.1. HgTe as interband material

In 2010, when I joined Guyot-Sionnest's group, there were already a few reports regarding the synthesis of HgTe QD, but they were all still below 3µm. For a review on the topic the reader should refer to references.[105] Briefly, the colloidal synthesis of HgX compounds started with aqueous precipitation[106-108] and evolved towards organic methods,[109-111] striving for increasingly better control of size and optical properties. In 2010, Sean Keuleyan developed the first synthesis of HgTe CQD[8] with a tunable band edge all over the 3-5 µm range (see Figure 10a). This early material was strongly aggregated (see Figure 10b) which leads to broad optical features but has the advantage of leading to strong interdot coupling and high mobility of the order of 1 $cm^2V^{-1}s^{-1}$, without ligand exchange. The synthesis was then further improved[10] to reduce the aggregation (see Figure 10d) and obtain sharper optical transitions (see Figure 10c). An even higher level of control was reached by Sean by the time I had left Chicago to obtain an improved control of the monodispersity[112] (see Figure 10f). The resulting spectrum now presents several excitonic transitions (see Figure 10e). The interband edge of HgTe can now be tuned from 3 to 12 µm depending on the CQD size (see Figure 9c). More recently, in collaboration with Sandrine Ithurria, we developed 2D NPL of HgTe and HgSe (see Figure 10g). Direct synthesis is not yet possible, so we use an indirect path where we first synthetize nanoplatelets (NPL) of CdTe or CdSe and then conduct a cation exchange process[113-115]. This amazing process consists in exchanging the cation while keeping the anion lattice unchanged. In the case of HgX NPL this was especially challenging because of the softness of Hg-based material. The trick was really to slow down the reaction by using bulky (i.e., complexed with long ligand) precursors of the mercury. So far they remain in the near-IR (800-900 nm), but they bring all the optical advantages of the 2D geometry (i.e., narrow optical feature and fast emission; see the next chapter for more details).

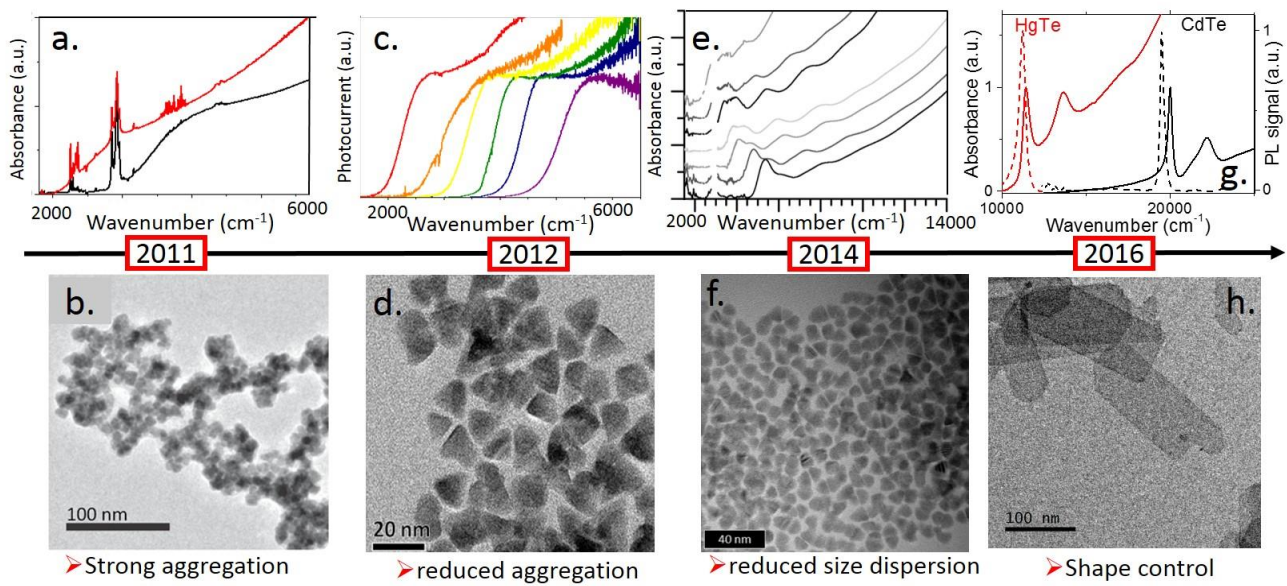

*Figure 10 Recent progress regarding the synthesis of HgTe nanocrystals. In 2011, agregated CQD were the first to present infrared absorption in the 3-5µm range (see ref 8). In 2012, an improved synthesis with reduced agregation was proposed (see ref 10) and led to sharper optical transistion. In 2014, improved monodispersity was obtained and led to the observation of several excitonic transition (see ref 112). In 2016, 2D nanocrystals of HgTe were obtained from cation exchange of CdTe NPL (see ref 31).*

### 2.2.2. HgSe as Intraband Material

- *Synthesis*

While HgTe behaves almost like an intrinsic semiconductor, the two other mercury chalcogenides HgS and HgSe present a very interesting behavior: they are self-doped. In the field of CQDs the doping remains an open challenge, and no systematic method has been identified[116,117] to obtain on-demand doping. The Guyot-Sionnest group has been leading the field in this direction and was the first to observe this self-doping[118,119]. We bring our contribution to the field of self-doped nanocrystal with the work of several students, notably Marion Scarafagio and Patrick Hease for the synthetic part, and Adrien Robin for the understanding of doping. We first proposed a new synthetic path for the synthesis of HgSe[26]. The synthesis is based on the reaction of a mercury oleate oleylamine complex with a selenium precursor. Depending on the targeted size of the CQD, two types of precursor can be used. Selenium complex with trioctylphosphine (TOPSe) allows the growth of small CQD in the 3 to 15 nm range (see Figure 11a). However, the presence of TOP, which strongly binds to mercury, prevents the growth of larger objects. So we switched to a phosphine free synthesis[120] and used $SeS_2$ as precursor. Particles up to 50 nm can be obtained (see Figure 11b). The size of the CQD is then finally controlled by tuning the temperature (from 70 to 120°C, above which the mercury complex is degraded) and reaction time (from 1 min to 30 min, typically).

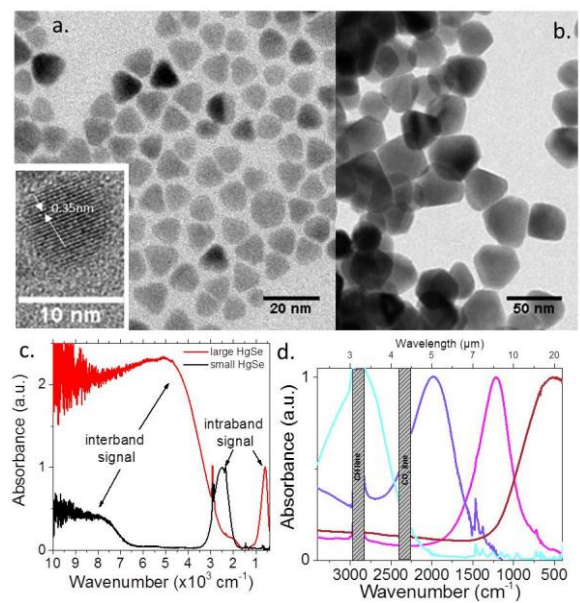

*Figure 11 a. and b. are TEM images of HgSe nanocrystals of different size. c. Absorption spectrum of two sizes of HgSe CQD. The spectrum presents two main contributions. At high energy, the interband transistion is observed, while at low energy a peak absorption results from intraband transistion within the conduction band. d. The energy of this intraband peak can be tuned all over te mid infrared thanks to size tunning, adapted from 26.*

The spectrum of HgSe present two distinct features. At high energy, a broad signal results from the interband absorption. In the mid-infrared, a peak signal is the signature of the intraband transition and the evidence for the doped character of these nanocrystals. The intraband peak can be tuned from 3 to 20 µm while tuning the CQD size from 4 nm to 50 nm (see Figure 9c and Figure 10d). This value of 20 µm, almost in the THz range, is the reddest reported for CQDs[26]. The presence of this intraband absorption is really opening new possibilities for quantum engineering based on nanocrystals

- *Self-doping origin*

The presence of this intraband band signal raises questions about the origin and possible control of the magnitude of the doping. This work has been the object of the end of Adrien Robin's PhD project. Before proposing a model which would rationalize the self-doping of HgX CQD, a few observations are worth mentioning. Rapidly after the synthesis of these materials, it appears that the intraband signal magnitude can be strongly tuned while tuning the surface chemistry. The growth of a CdS shell or a ligand exchange step strongly tunes the ratio of the inter- to intraband absorption (see Figure 12a-b). We can quantify this change of absorption in terms of the population of the 1s state. Typically the carrier density is reduced by a factor 10 as we switch from the initial capping ligand toward $S^{2-}$. This change of the population while tuning the size is more important for smaller CQDs and strongly depends on the ligand. As we look for a doping within the bulk of the CQDs, we have to consider the high working function of HgX, which is around 6 eV. This makes the reduction of the QD by water/environment possible according to the reaction $2QD + H_2O \longrightarrow 2QD^- + 1/2 O_2 + 2H^+$. Thus the stable form for HgS and HgSe is the negatively charged CQD, while HgTe which is closer to the vacuum level is not reduced and remains neutral. This mechanism was the further confirmed by electrochemistry.[121] The effect of a surface chemistry modification is to add dipoles. This leads to a shift of the vacuum level according to the equation. The anionic ligands point their negative charge toward the CQDs, which screen this field by bringing positive charge at the surface, which tends to bend the bands upwards and reduce the doping (see Figure 12d). The equation $\Delta E_{vac} = -N \frac{\mu_\perp (ligand)}{\varepsilon_0 \varepsilon_{ligand}}$ with $N$ the dipole surface density, $\mu$ the dipole magnitude, $\varepsilon_0$ the vacuum permittivity and $\varepsilon_{ligand}$ the medium dielectric constant allows proposing the Figure 12e for the dipole dependence of the band structure as dipoles are added. Small CQDs, which are confined, have a conduction band above the $O_2/H_2O$ couple and are not reduced as observed. Then the larger the CQD, the lower the confinement and the larger the reduction (i.e., the higher the doping). The effect of dipole is then to bring a surface tunability to the bulk reduction of the CQDs. This is an interesting approach because we reach tunability in the 0.1 to 2 electron(s) per CQD which is not so easy to obtain. Finally, we also proposed a model to relate the population of the 1s states to the magnitude of the dipole, which leads to an optical measurement of the dipole intensity (see Figure 12c).

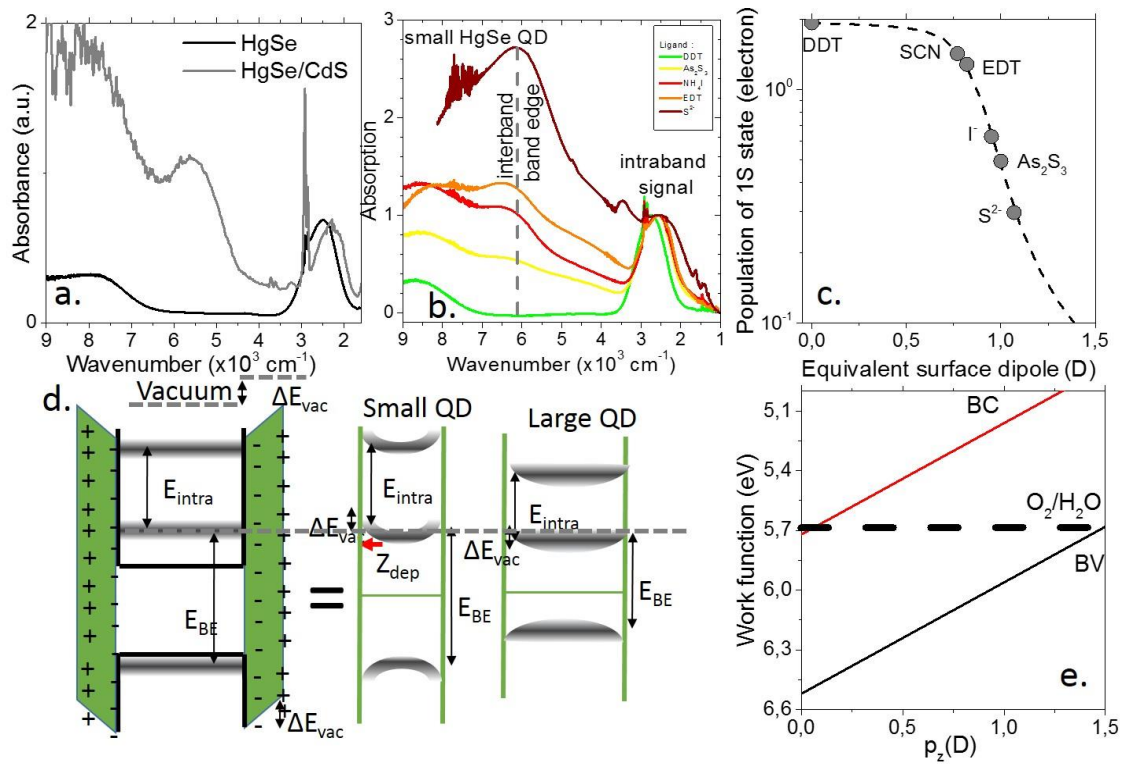

*Figure 12 a. Absorption spectrum of HgSe and HgSe/CdS CQD. Once the shell is grown the $1S_h$-$1S_e$ transistion appears and the relative magnitude of the intraband transistion is reduced. b. Absorption spectrum of HgSe CQD capped by different ligands. c. Population of the 1s state as a function of the surface dipole for HgSe CQD capped with different dipole. d. Scheme of the electroic structire of HgSe CQD functionalized by surface dipole which bend the band on the surface. e. Energy of the conduction and valence band as a function of the surface dipole magnitude, adapted from 35.*

### 2.3. Electronic transport in narrow band gap nanocrystals

I now would like to discuss the transport properties of the HgX CQDs. HgX CQDs are not the only colloidal nanocrystals with absorption in the mid-infrared. There are a few reports of doped oxide[122,123] and doped silicon nanocrystals[124] which show mid-IR plasmonic features. However, what makes HgX CQDs appealing is that not only do they absorb, but they also give a photoconductive signal. This is why they will become the next generation of low-cost infrared photodetectors. For this research, I was supported by several students: Sean Keuleyan at U. Chicago, Marion Scarafagio, Adrien Robin at ESPCI and Clément Livache at INSP.

### 2.3.1. Tuning the surface chemistry to boost the carrier mobility

- *Strategies to increase the local coupling*

To make a film of CQDs conductive, the general route is described in Figure 13.[125] The idea is to reduce the height and length of the tunnel barrier in order to increase the interdot coupling. This can be obtained from solid state ligand exchange (i.e., the film is already formed) or in liquid phase. We first investigated the classic ligand exchange with ethandithiol (EDT). EDT is a two carbon chain with a thiol function at each end. The dithiol crosslinked the CQD, which not only leads to a higher electronic coupling but also mechanically hardens the film. With EDT treatment, we indeed observed a rise of the conductance, but we also observed a non-monotonic temperature dependence of the current.[14] This effect actually results from an oxidation of the CQD surface and can be avoided by preparing the film in air-free condition (i.e., in a glove box).

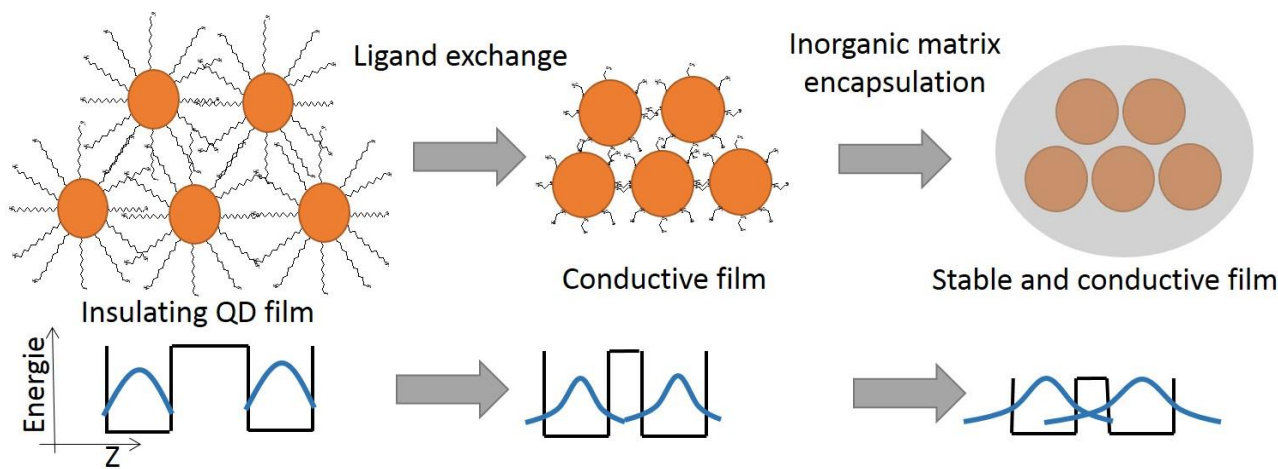

*Figure 13 Scheme of the strategy to make a CQD film conductive. At first CQD are capped with long capping ligands which is analog to two wells coupled by a large and high tunnel barrier. As ligand exchange toward short ligand is conducted, the tunnel barrier height is reduced. If all the organic ligands is removed, not only it is possible to play on the barrier length but also on its height.*

However, this approach does not address one key difficulty which occurs in the IR and results from the ligand absorption in the IR. In particular the C-H bond has a strong resonance at 3000 cm$^{-1}$. If the absorption of the CQD is close to this value, an energy transfer from the CQD to the ligands can occur. In fact, because of the vicinity of the two materials, this transfer is extremely efficient[126] (>99%) and this quenches the PL and photodetection efficiency. For this reason, removing the organic ligands is even more important for IR materials.

- *$As_2S_3$ a well suited surface chemistry for infrared*

We investigated an inorganic approach for the HgTe CQD capping. This inorganic matrix has to be (*i*) IR transparent, (*ii*) form a type I heterostructure with the CQD (i.e., the CQD has to remain the active IR material) and leads to (*iii*) a reasonable mobility. We identified $As_2S_3$ as an interesting candidate[127,128]. In a first step $As_2S_3$ powder is dissolved in propylamine. This solution is then diluted in ethanol. The HgTe CQD film is dipped in this solution and then rinsed to remove the excess of ligands. The deposition process and the ligand exchange are repeated at least a second time to fill the cracks (see the SEM image on the bottom right part of Figure 6) resulting from the film volume reduction occurring as long ligands are replaced by short ligands. A top capping layer is added to insulate the CQD film from the environment (see Figure 14a).

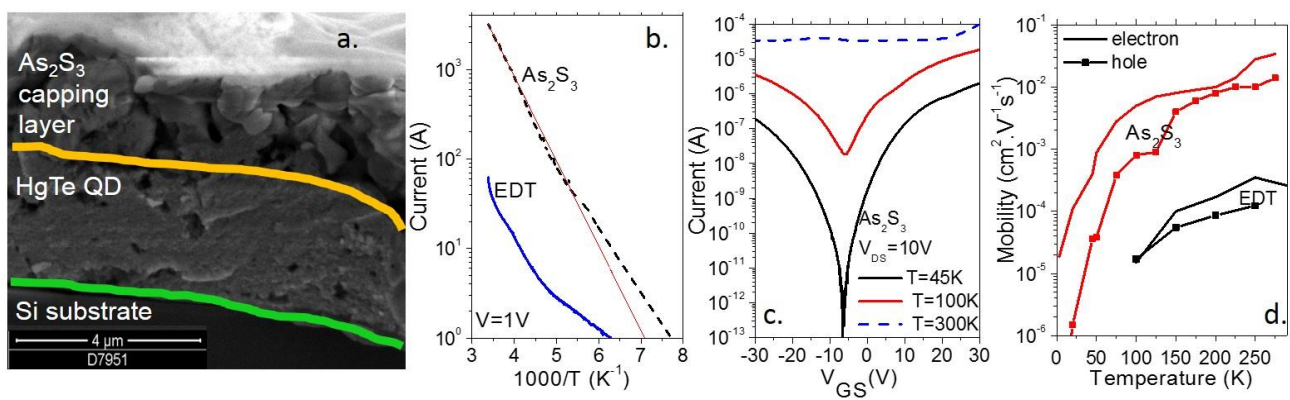

*Figure 14a. SEM image of HgTe/ As$_2$S$_3$ hybrid structure. b. Temperature dependence of the current for As$_2$S$_3$ and EDT capping of air free proceseed HgSe CQD thin film. c. Transfer curve of HgTe CQD capped with As$_2$S$_3$ thin film transistor at different temperatures. d. Field effect mobility as a function of temperature for short organic ligand (EDT) and As$_2$S$_3$ caping ligand of HgTe CQD, adapted from 14.*

HgTe CQDs, prepared in air-free conditions, present an ambipolar behavior with an electron and hole mobility which are quite close (see Figure 14c). This situation is quite favorable for fast phototransport. In addition, the thermal dependence, which is very important for photodetection, presents an activation energy which is close to half the band edge energy.[14] This value maximizes the benefit of cooling by greatly reducing the dark current (see Figure 14b). It is also worth noting the improvement brought by the As$_2$S$_3$ ligands. First, it leads to higher activation energy (see Figure 14b) compared to EDT ligands. Secondly, the obtained mobility is two orders of magnitude higher[14] (see Figure 14d).

We re-explored the surface chemistry of HgSe during the internship of Marion Scarafagio. We conducted the ligand exchange in liquid phase to maximize the removing of organic ligands. However, instead of conducting the measurements with a simple back gate of SiO$_2$, we built a dual gate gating,[26] where the SiO$_2$ back gate is combined with a top electrolytic gate (see a scheme of the device in Figure 15a). This type of gate will be further discussed in the next chapter. The transfer curve is characteristic of a n-type material[26] (see Figure 15b). This confirms the doped character of the material and that doping is due to electrons. We have obtained very large mobility, up to 90 cm$^2$V$^{-1}$s$^{-1}$. This rise of the conductance results from the improved character of the ligand exchange. We also note the non-monotonic character of the current as a function of the CQD filling (see Figure 15b). Above 1 electron per CQD, some CQD are filled with 2 electrons on their 1s state. With only two states available, the Pauli blockage prevents an additional filling. Some CQDs become ineffective for transport until the p-states start getting filled.[26]

To conclude this discussion about transport, I would like to mention some recent work of our collaborator, the Aubin's group[24,129]. They have used our HgSe CQD in an on-chip tunnel spectroscopy configuration. Briefly, they (electro)spray the particles on a chip where nanogaps have been prepared by e-beam lithography. Once a nanoparticle is trapped between the electrodes, the current overcomes a threshold and the chip is transferred to a cryostat. They obtain a tunnel map for this HgSe CQD (see Figure 15c).[37] In particular, the map clearly exhibits two band gap energies as a function of the gate bias (i.e., CQD filling): a large gap, due to the interband transition, and a smaller one, due to the intraband transition, once the Fermi level reaches the 1s state. This demonstrates that doping exists at the single particle level and brings one more piece of evidence to the self-doped character of these CQDs.

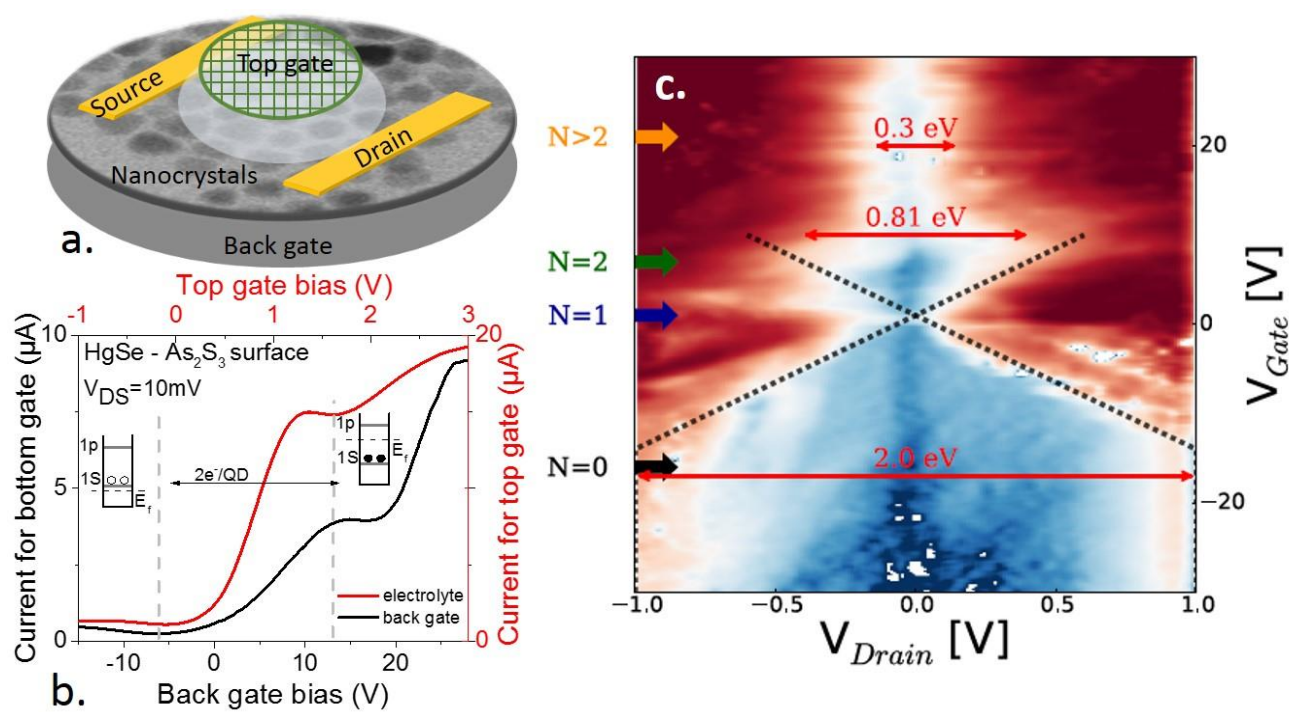

*Figure 15 a. Scheme of a field effect transistor, which channel is made of HgSe CQD. b. Transfer curve (drain current vs gate bias) for a film of HgSe CQD. The curve is characteristic of a n type semiconductor. The local minimum result from the mobility edge once the 1s state is filled, adapted from 26. c. Tunnel map (conductance vs drain and gate bias) for a single HgSe CQD. Two gaps are clearly observed, adapted from 37.*

### 2.3.2. Dynamic aspect of transport

On the subject of transport, a last aspect needs to be discussed, and concerns noise. This frequency-resolved contribution of the current is particularly critical for infrared photodetection. Indeed, for IR detection, the signal to noise ratio tends to be low due to thermal activation through the narrow gap. The noise needs to be measured to accurately estimate the detectivity (i.e., the signal to noise ratio) of a detector. We were the first to investigate experimentally the noise in CQD solids.[9,18,130]

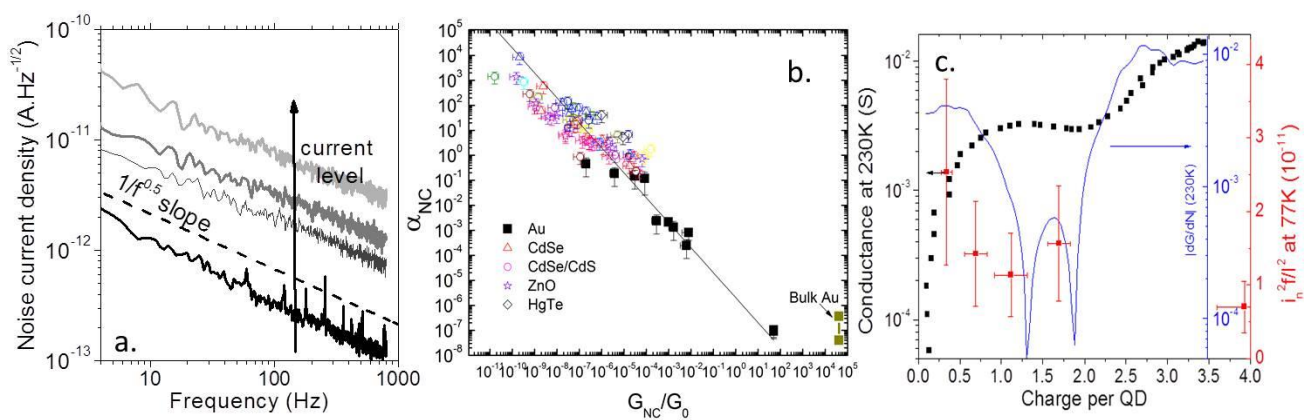

*Figure 16 a. Noise current density as a function of signal frequency for CdSe NPL film under different biases. b. Hooge constant normalized by the number of nanocrystals as a function of the interparticle conductance for a broad range of material and nanoparticle sizes. Conductance, transconductance and noise intensity as a function of the CQD carrier density, from ref 18.*

So far our observation from a broad range of materials (metal, wide band gap and narrow band gap semiconductors) is that at low frequency (<kHz), the noise is always limited by a *1/f* noise (see Figure 16a). The exact exponent from the frequency dependence might fluctuate a bit, but values between -0.8 and -1 include most of the measurements. Very little is as yet understood about the origin of the *1/f* noise,[131,132] but Hooge's law[133] $i_n^2 = \frac{\alpha \cdot I^2}{N \cdot f}$ is the most widely-accepted empirical model. It relates the noise spectral density $i_n$ with the current magnitude *I*, the number of carriers *N* involved in transport and the frequency *f*. α is a constant that Hooge spent his life to measure, and is equal in metal film to $10^{-4}$-$10^{-3}$. In a CQD film, not only *1/f* noise prevails but its relative magnitude is very high; α around 1 is typical.[9] We came to the point where understanding more about the origin of the noise was necessary. A first key question was whether the noise results from charge or mobility fluctuation. To answer this question, we split the noise in two components according to the equation $\frac{i_n^2}{I^2} \prec \left(\frac{\partial I}{\partial n}\right)^2 \frac{\delta n^2}{I^2} + \left(\frac{\partial I}{\partial \mu}\right)^2 \frac{\delta \mu^2}{I^2}$. Then we used the non-monotonic behavior of the transfer curve as a function of charging. I mentioned previously that the conductance as a function of the CQD filling presents a local maximum when 1 electron per dot is reached, and a local minimum when 2 electrons per CQD is reached (see Figure 15b or Figure 16c). At these two points we can cancel the transconductance $\frac{\partial I}{\partial n}$. We then follow the noise as a function of the CQD filling. If the carrier fluctuations were prevailing, we would have expected a minimum of noise in the 1 to 2 electron per CQD range. We did not observe such a minimum and we concluded that noise most likely results from mobility fluctuations themselves, resulting from the distribution of distance between the CQDs.

Finally, we have identified a general law to relate the noise magnitude to the inter-CQD conductance. This rule applies for a broad range of materials (Au, ZnO, CdSe, HgTe) (see Figure 16b). This power dependence of the noise with the conductance implies that parameters such as the nanoparticle size, or even the material composition, are not driving the noise magnitude. On the other hand, when the coupling gets strong it is possible to recover noise levels similar to the bulk.

### 2.4. Toward a new generation of infrared detector
#### 2.4.1. State of the art

During my research, I have dedicated significant effort to the measurement of device performance. This means that we systematically measure the calibrated photoresponse, the time response, the noise, and finally, we can evaluate the detectivity. In terms of photoresponse the responsivity of HgTe CQD-based devices typically ranges from a few mA.W$^{-1}$ to 100 mA.W$^{-1}$. This corresponds to quantum efficiency around 10% (see Figure 17a). The time response of HgTe can be fast with a cut-off frequency above 100 kHz. With HgSe, the responsivity can be larger, up to 600 mA.W$^{-1}$, but at the price of a much slower photoresponse. The cut-off frequency drops around 50 Hz. The main drawback relative to intraband-based devices remains their large dark current which leads to smaller detectivity around $10^8$ jones at room temperature[26] and $10^9$ at liquid nitrogen temperature[119]. With HgTe the dark current can be lower, and thanks to higher activation energy can be reduced further at low temperature. The best devices are currently based on HgTe (see Figure 17b) with detectivity above $10^{10}$ jones. The best mid-IR device has been proposed by the Guyot-Sionnest group with a cut-off wavelength at 5µm in a photovoltaic mode[134]. Comparison of the obtained performances with existing technologies shows that for monopixels the CQD-based device already offers a competitive level of performance (see Figure 17b). It is also worth mentioning that the U. Chicago groups generated the first image by coupling a film of HgTe CQDs to a bolometer readout circuit.[135,136]

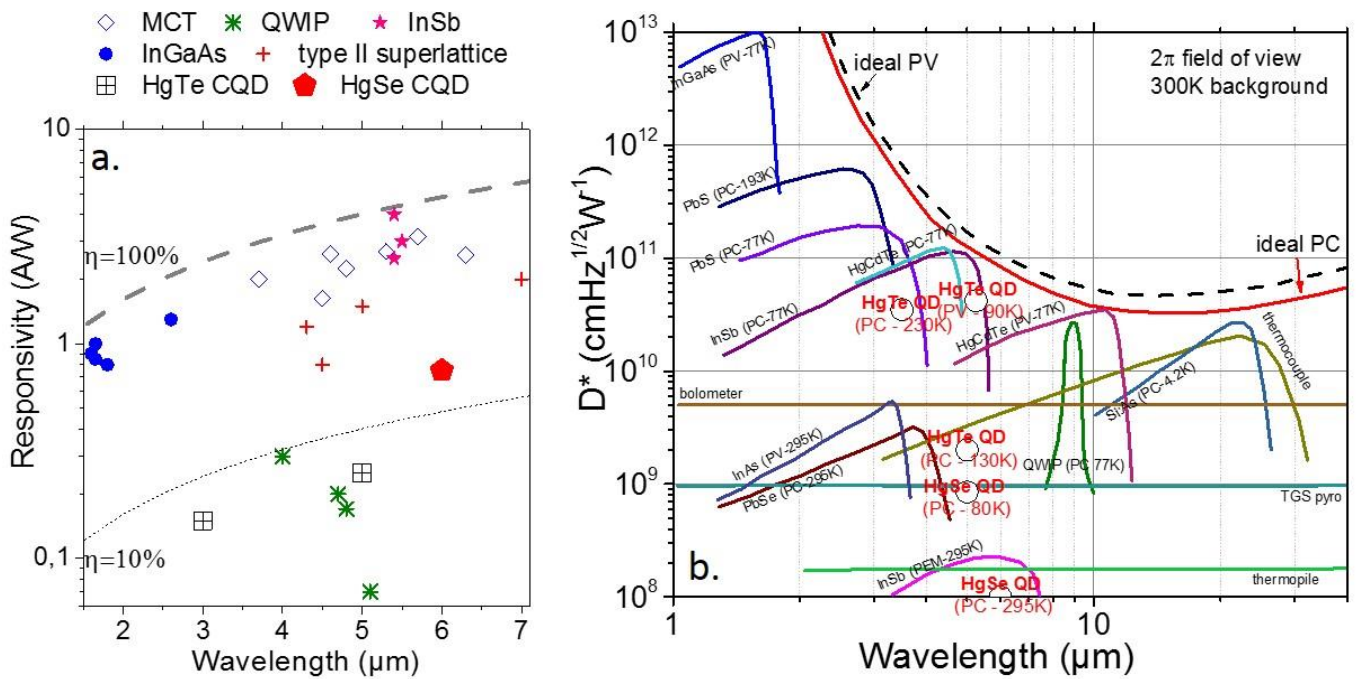

*Figure 17 a. Responsivity as a function of the wavelength for different technologies of infrared detectors, adapted from 26. b. Detectivity as a function of wavelength for different technologies of infrared detector. The point of HgX CQD are given in red, adapted from 39.*

### *2.4.2. Current limitations*

I have identified the following limitations to current CQD-based mid IR detectors

- **Control of the doping is a key issue**. This is true for infrared detection but also true for the design of photovoltaic solar cells. I want to address this problem both from the material point of view and on the device side. Methods to control the doping level of the material and control it *a posteriori* using a gate need to be developed. Doping will also be critical for the device in photovoltaic geometry to achieve an efficient charge dissociation.

- **We still lack data about the electronic structure of the HgX CQDs**. As a result we do not know where are the electronic levels and the design of devices is performed by an trail and error approach. We need more information about the absolute energy levels in HgX CQD. Too many material parameters remain unknown or are determined with a very poor accuracy (0.1 eV and more for band edge energy of 0.1 to 0.5 eV). A very common methodology consists of using bulk parameters, but this approach is clearly a very rough estimation. I want to go beyond this approach and **systematically probe the electronic structure of the HgX CQDs** using photoemission and IR spectroscopy.

- **The noise in CQD solids** comes under a *1/f* noise and its relative magnitude is important (see ref 18). The noise raises the question of how to reduce the dark current.

- The last aspect relates to **the dynamics of this system**. Again, this is an open question. From an applied point of view, we do not know how fast an IR CQD-based device can operate. Is there any fundamental limitation, or are the limitations related to the material?

- The **stability of the device** remains also an open question. The device performance needs to remain stable over a long timescale and no measurement has been conducted to determine the device lifetime. Certainly the encapsulation of the device within a matrix will be necessary to reach an appropriate level of stability.

# 3. 2D NANOCRYSTALS

> Students involved in this work : Adrien Robin, Daniel Thomas, Remi Castain.
>
> Publications relative to this works: 16, 17, 19, 20, 23, 25, 28.
>
> Reviews related to this work: 22, 30.
>
> Conference proceedings relative to this work: 48, 49.
>
> Patents relative to this work: 53-56.

## 3.1. Cadmium chalcogenide nanoplatelets
### 3.1.1. Materials

Of the material challenges in the field of nanocrystals, control of the shape and dimensionality has received a tremendous amount of effort. The Alivisatos group first reported the synthesis of non-spherical objects with rod shape[137]. It was only later that flat objects with 2D dimentionality were obtained. It was first proposed by the Hyeon group[138,139] and then improved by Ithurria *et al,*[140,141] although there were already reports concerning flat objects in particular by the Schaak group.[142-144] What makes the cadmium chalcogenides fairly special is the unique control of the thickness that has been achieved. So far, the optical properties of CQD are generally limited by the inhomogeneous broadening (i.e., size distribution) of the optical features. For example, CdSe CQDs with emission around 550 nm have a FWHM of 25 nm for good synthesis. With CdX nanoplatelets (NPL), the FWHM can be below 10 nm[22,30,145]. This corresponds to a broadening between $1-2k_bT$. CdX NPL are 2D objects with a lateral dimension above the Bohr radius and a thickness of a few monolayers (1-2 nm). Only the thickness direction is confined, and this direction is perfectly flat (i.e., no roughness).

At first sight, the growth in solution of anisotropic objects is not obvious, as this is a medium where random isotropic Brownian forces are applied. We discuss the possible mechanisms which are likely to induce anisotropic 2D growth in a recent review.[30] Four main mechanisms have been identified:[30](i) The atomic lattice is it-self anisotropic. This is typically the case of wurtzite NPL for which the growth occurs preferably along the c axis. (ii) Anisotropy may be induced by lattice defects. A third way (iii) to induce anisotropy results from ligand engineering. Ligands might form a lamellar template which favors growth within the plane. Finally, (iv) the anisotropy might be reduced by the self-assembly of pre-formed, small isotropic nanocrystal which connect through specific facets. This mechanism is very common for lead chalcogenides[146]. In the case of zinc blend CdX NPL, such as the one we will investigate in what follows, the exact mechanism is still under debate,[147] but the ligands and their relative solubility play a key role.

### 3.1.2. An original approach for cleaning and film deposition: electrophoresis

One of my first steps in the field of 2D NPL was actually focused on the development of alternative methods to deposit film of NPL using electrophoresis. With Patrick Hease, we demonstrated that it was possible, and additionally that the method can be used to sort NPL from the side products of synthesis. The synthesis of CdSe NPL relies on the initial growth of CdSe small seeds. To do so, cadmium complexed with long carboxylate chains (myristate or oleate) reacts with selenium powder. As the growth step starts, additional cadmiums complexed with short carboxylate chains (acetate) are introduced. This reduces the solubility of the seed and leads to the anisotropic growth of NPL. Nevertheless, spherical CQD comes as side product of the reaction. This typically illustrates the absorption spectrum of the CdSe NPL. The crude mixture of the reaction presents some absorption below the first excitonic peak of the NPL which comes from the 0D CQD (see Figure 18a). To get rid of them, selective precipitations are performed. We use the difference of colloidal stability between the NPL (poorly stable) and the CQD (highly stable) to split the two populations. However, this procedure is time and solvent consuming. Thus during Patrick Hease's tenure as technician, we developed an alternative method based on electrophoresis[20].

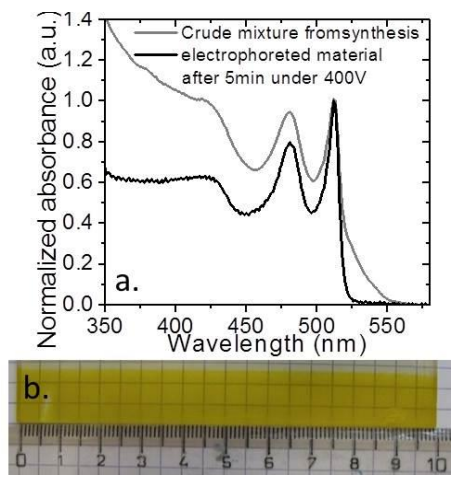

*Figure 18 a. Absorption spectrum of the crude mixture resulting from the synthesis of 2D NPL before and after their selective cleaning. b. Image of a electrophoretically deposited fikm of CdSe NPL. The scale is in cm, adapted from ref 20.*

We first investigated the electrophoretic deposition of the model solution of pure NPL and CQD of similar size to the side product obtained during synthesis. The TEM image and absorption spectrum associated with each material are shown in Figure 19a-d. Some acetone was added to a solution of nanoparticles dispersed in their non-polar solvent, to destabilize the solution. We then dipped two FTO coated glass slides used as electrodes in the solution. We applied a 400 V bias ($F≈400$ V.cm$^{-1}$) between them. We observed a deposition of material only on the positive electrode. The deposition takes less than 1 min in the case of NPL, while more than 1 h is necessary in the case of CQD (see Figure 19f). We then further used this difference of kinetic deposition to sort some small amount of NPL diluted in a large amount of CQD. From this mixture, we were indeed able to extract only the NPL (see Figure 19e). The selectivity of the process has been estimated to be in the 100-300 range (see Figure 19g). We finally applied the method to sort the NPL from their crude synthesis mixture. This process is far more challenging because of the complexity of the mixture—the solvent is more viscous and there are some unreacted species. In spite of these difficulties we obtained on the positive electrode a film made of mostly NPL. The deposition is homogenous over large scale (see Figure 18b). The obtained film can also be redispersed in good solvent if necessary.

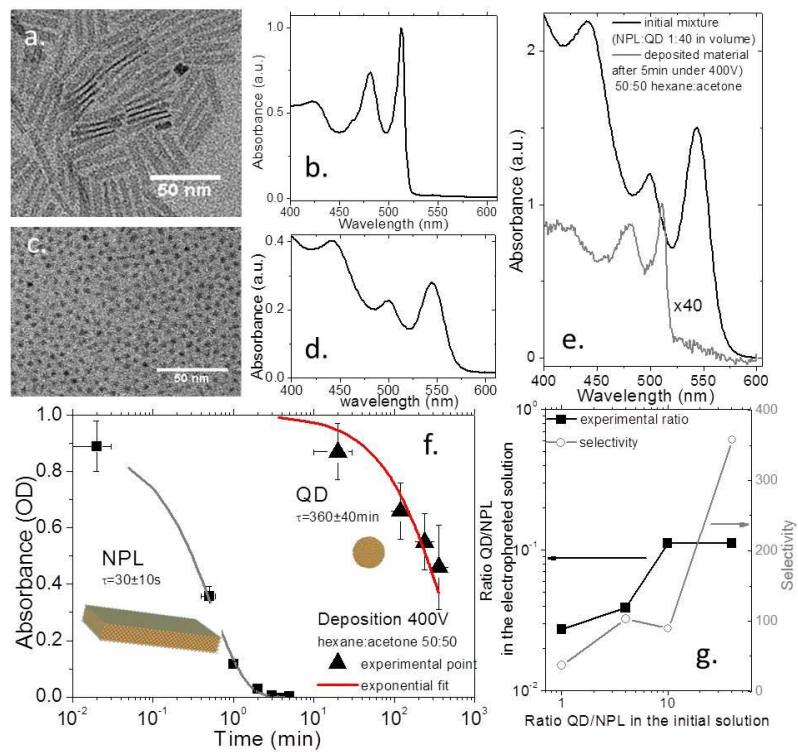

*Figure 19 a. TEM image of CdSe NPL. b. Absorption spectrum of CdSe NPL. c. TEM image of CdSe CQD. d. Absorption spectrum of CdSe CQD. e. Absorption spectrum of a mixture made of 40 equivalent of CQD and 1 equivalent of NPL. The absroption spectrum of the deposited material (grey line), similar to the spectrum of pure CdSe is also shown. f. Decay curve of the absorbance of a solution of CdSe NPL and CdSe CQD as a function of time during the electrophoretic procedure. g. Ratio of deposited NPL/CQD as a function of the ratio of NPL/CQD introduced in the initial solution.*

### 3.1.3. Impact of the dimentionality

The use of NPL instead of CQD for the design of optoelectronic devices raises the question of why should we expect different properties from these 2D objects compared with 0D CQD. Anisotropic objects raise great hopes for transport since we can expect a reduced number of hopping steps to reach the electrodes, and consequently there will be fewer tunnel barriers to go through. In addition, 2D platelets which can easily stack on each other should have a stronger wavefunction overlap compared to 0D CQD. On the optical side a well-defined optical feature with sharp transitions is obviously one of the key advantages of 2D NPL, even if it actually results from the growth mechanism rather than from the dimensionality itself. In addition, their short PL lifetime makes them promising for light emission and is certainly responsible for the reduced laser threshold obtained with NPL.[148,149]

A very striking difference between NPL and 0D CQD is their dielectric confinement. This effect is also a key difference between the 2D colloidal objects and those obtained by epitaxy. For epitaxially grown heterostructures, the change of the dielectric constant from one material to the other is very small, while the change of the dielectric constant is great between the semiconductor and its solvent. As a result, mirror charge effects are boosted as we switch from 0D to 2D shape and charges are pushed way from the surface. However, this effect is not observed on the electronic spectrum, because it is mostly balanced by the binding energy of the electron hole pair. As we can see later this will have a severe impact on transport properties. In particular, the binding energy is quite large for these NPL. Its value has been estimated to be 200-300 meV.[150,151] This magnitude has to be compared with the few meV to few tens of meV occurring in 0D CQD. This makes the NPL quite close to what can be observed on transition metal dichalcogenides (TMDC) for which a binding energy of 0.5-1 eV is quite typical.[152] Managing the electron hole pair dissociation will be crucial for the design of photodetector

## 3.2. Transport in nanoplatelet arrays
### 3.2.1. Electrolytic Transistor

As a first step in the use of NPL for optoelectronics we have decided to design a field effect transistor (FET). After a few unsuccessful attempts for conventional back gate FET, we turned our focus to electrolytic transistors. A key motivation for electrolyte gating comes from the very large gate capacitance. In a conventional dielectric transistor the gate capacitance is given by the ratio of the dielectric constant divided by the dielectric thickness. For a 400 nm $SiO_2$ layer, this corresponds to a gate value of 10 nFcm$^{-2}$. With electrolytes this is the length of the ionic layer which drives the capacitance. As a result, gate values above 1 µF.cm$^{-2}$ are commonly obtained.

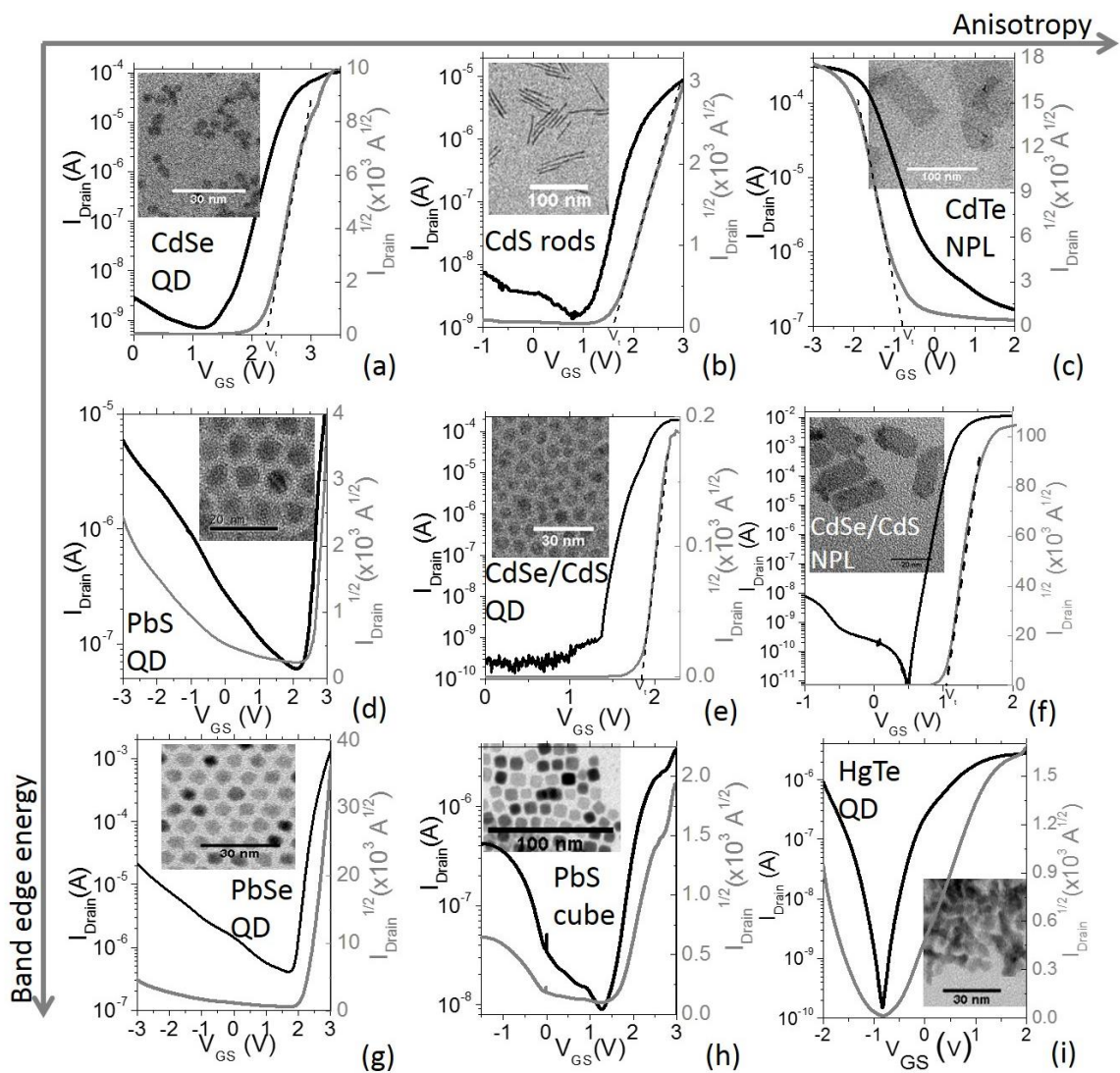

Figure 20 Transfer curve (drain curret vs gate bias) for broad range of CQD materials.

Electrochemical FET have been widely investigated by the Guyot-sionnest group[153] and Vanmaekelbergh group[154]. However, they were using liquid electrolytes which suffer from two main drawbacks. They have to be prepared and used in air-free conditions and the electrochemical cell remains bulky/leaky. Instead, we developed a very simple ion gel gating[155] which has become the most used electrical characterization of our materials. To outline the process briefly: we combine a ligand exchanged film of NPL with ion gel prepared of LiClO$_4$ dissolved in polyethylene glycol (PEG). PEG is already a solvent of the Li salt and no other solvent has to be used. This prevents the exposure of the CQD to an annealing step in the presence of non-solvent.

Employing this approach we obtained FET with strong current modulation (see Figure 20). The method is versatile and can be applied to a broad range of material, from that with wide band gap (CdSe, CdS, CdTe) to narrow band gap (PbS, PbSe and HgTe)  and with various shapes. We observed that wide band gap materials are generally unipolar. Indeed, with the exception of CdTe, they are almost all n-type. On the other hand, narrow band gap materials are ambipolars. This observation might be in contradiction with previous results in the literature, but it actually results from the more efficient gate compared to dielectric gating that we used. We can push the charge carrier density further for both (i.e. electron and hole) carriers. This electrolyte gating allows low bias operation and we obtain on state while the gate bias remains below 3 V, which is generally the range of stability of the electrolyte.

We then investigated more deeply the gating process on its own. First we confirmed that charges are indeed injected in the quantum states of the NPL. To do so we conducted spectroelectrochemistry experiments, in which we look at the bleach of the absorption while electrons are injected in the conduction band (see Figure 21a). This confirms that charges are not injected only in surface traps. As for liquid electrolyte, ion gel allows the charging of thick film (500nm have been successfully charged). We indeed see that the injected charge increases almost linearly with the film thickness (see Figure 21b). This is made possible because ions percolate between the nanocrystals.

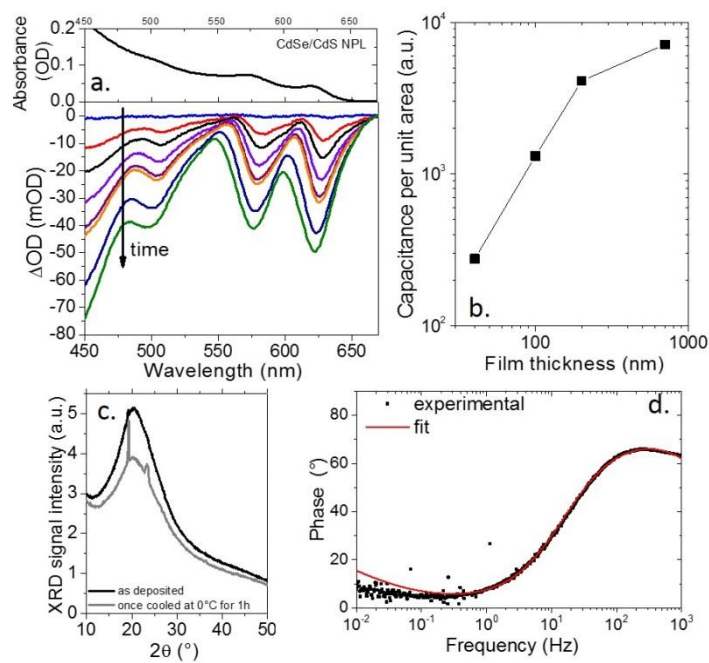

*Figure 21 a. (top) Absorbance spetrum of CdSe/CdS NPL film. (bottom) Change in the absorption of a CdSe/CdS NPL film as a function of time while a 2V gate bias is applied. b. Injected charge in a film of CdSe/CdS NPL as a function of the film thickness. c. X-ray diffraction pattern of LiClO$_4$/PEG electrolyte as prepared and once cooled below 0°C. d. Phase relative to the excitation of the electrolytic gate as a function of the signal frequency.*

During the postdoctoral research of Remi Castaing, we investigated the temperature and frequency dependence of ion gel gating. We observed that below 0°C, the gating properties are lost. We can relate this loss of ion mobility to a change of phase within the PEG. At high temperature, the PEG forms an ether crown which solvates the Li$^+$ ions. As the ions move the ether crown reorganizes. That is why at high temperature we observed an amorphous signal (a broad peak, see Figure 21c) for the electrolyte. Below 0°C, the electrolyte starts to freeze, as shown by the appearance of narrow diffraction peaks in the diffractogram, and the ether crown can no longer reorganize. This is a current key limitation of the electrolyte, which is only tunable at a high temperature.

Regarding the impedance measurements obtained on the ion gel gating, we can distinguish two regimes[156] (see Figure 21d). At high frequency, the phase is close to 90°, which is typical of capacitive behavior. In this regime, a double layer is formed, just on surface. To achieve the bulk charging of the CQD film, the ions need more time to percolate and this diffusion process comes with a decrease of the phase signal close to 0°. Recently, it was even proposed by Puntambekar *et al*[157] that with $LiClO_4$, not only there is percolation of the $Li^+$ between the nanocrystals ions, but in presence of small counterions $Li^+$ can even intercalate within the atomic lattice of the CQD. This intercalation is associated with slow dynamic and might be responsible for the more ohmic behavior.

### 3.2.2. From photoconductor to phototransistor

In a second step of this aspect of the research, we used the electrolytic transistor for its gate tunable properties. The idea was no longer to use the transistor as a probe of the carrier density but rather as a way to prepare the system to a certain operating point. This work was motivated by our preliminary work on CdTe[16] and CdSe NPL film. NPL films can present a strong current modulation under illumination (three four orders of magnitude) but the overall photoresponse is weak in the 10µA.W$^{-1}$. Given the low dark current in this system, resulting from the wide gap nature of CdX compounds, we can "sacrifice" a bit of the modulation (*i.e.* increase the dark current) to boost the photoresponse. In a photoconductive mode, the photocurrent typically scales like the light power. On the other hand, with a transistor operating in its subthreshold regime, the current presents an exponential dependence with the carrier density.

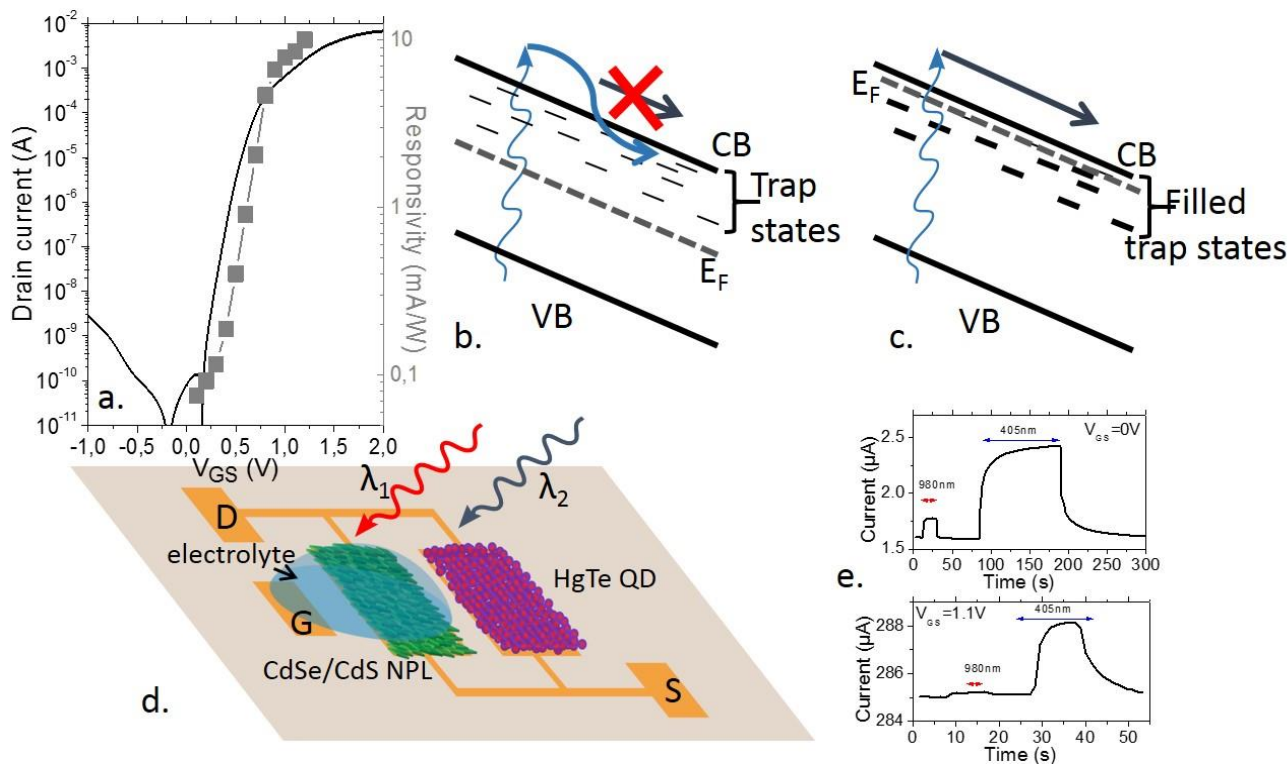

*Figure 22 a Drain current and responsivity of a CdSe/CdS NPL film as a function of the applied gate bias. b. Scheme of the band structure of CdSe/CdS NPL in presence of electronic traps which limit the electron lifetime. Once the electronic traps are filled due to a shift of the Fermi level (part c) the time spent by the electron within the conduction band is extended. d. Sheme of a dual color*

*photodetector, which photoresponse can be tuned from visible only to visible + infrared thanks to a gate bias, from ref 19. e change of the photoresponse under visible (405 nm) and infrared (980 nm) of the device of part d, for two gate biases.*

Our idea was to apply a gate bias on the CdSe/CdS film to bring it just below the turn on point. Then the excess of photogenerated carriers are expected to generate more photocurrent. We indeed observed a gate dependent photoresponse (see Figure 22a). Thanks to the gate we can increase the photocurrent by 2-3 orders of magnitude. The mechanism to explain this increase is called photogating and is actually a very common mechanism in the field of nanocrystal-based photodetection. This same mechanism also drives the photocurrent in the CQD graphene hybrid—see the section dedicated to the Van der Waals heterostructure. The CdSe NPL are n type materials,[17,25] as a result the electron can flow while the hole gets trapped. Gain is generated because the electron can recirculate several times during the lifetime of the hole. The latter is actually limited by external recombination with the environment and possibly a bimolecular process (*i.e.* a recombination with an electron). To maximize the gain, also defined as the ratio of the minority carrier lifetime over the transit time, it is critical that the electron spend its time to drift in the conduction. However, the electron can also get trapped (see Figure 22b). By applying a positive gate bias, the electron traps get filled and we observe an increase of the electron lifetime.

This strategy to control the photoresponse of a system through the use of gate bias was then further used for the design of a bicolor detector. We coupled a film of HgTe CQD which has a cut-off wavelength in the IR with a film of CdSe/CdS NPL which only presents a visible absorption (see Figure 22d). The two films are connected in parallel and a gate is controlling the (photo) conductance of the CdSe/CdS NPL film. Without a gate the conductance of the narrow band gap material prevails and we observe a photoresponse both in the visible and the IR (see the top part of Figure 22e). With the gate bias, the film of NPL is made sufficiently conductive to make its conductance prevail over that of the HgTe film, and we observe photoresponse only in the visible (see the bottom part of Figure 22e).

### *3.2.3. Transport at the single nanoplatelets scale*

Since most of the difficulty relative to transport in CQD solids results from the hopping transport, we investigated a strategy which removes it and takes advantage of the anisotropic shape of the NPL. One key issue which was not addressed by the electrolytic transistor is the large binding energy of the NPL. The applied energy drop per NPL remains below the exciton binding energy and the exciton dissociation is incomplete. For these two reasons, we chose to shrink the size of the device down to the single particle level.

We used a technology developed by J. F. Dayen at IPCMS (U. Strasbourg) to build a nanotrench, which is composed of two electrodes spaced by 20 to 100nm. This size typically corresponds to the lateral size of the NPL. The process only requires two steps of optical lithography in spite of the subwavelength character of the final device. The process involved a tilted evaporation which allowed us to overcome the diffraction limit. Details of the fabrication process are shown in Figure 23. The process in particular allows a large aspect ratio which is important to preserve absorption but is not so easy to obtain using e-beam, for example. A SEM image of the obtained device before and after its functionalization by NPL are shown in Figure 23c-d.

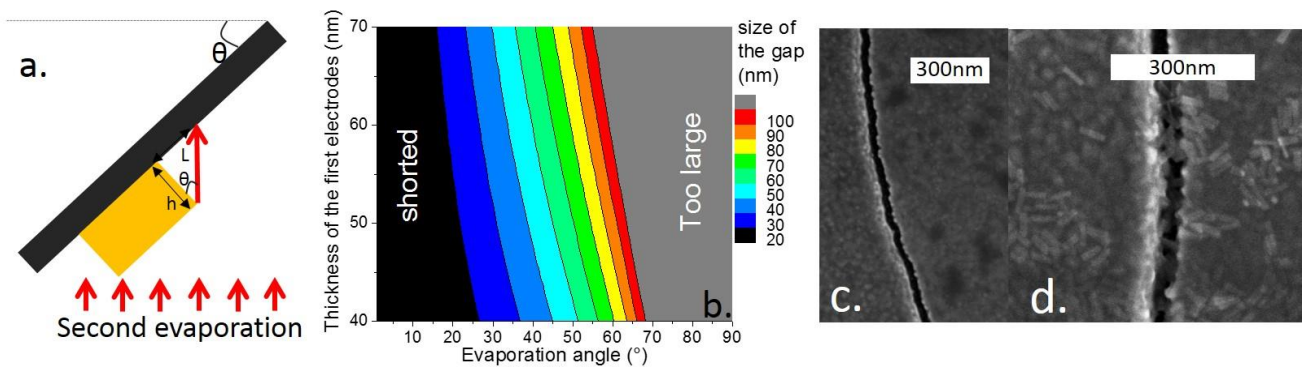

*Figure 23 a. Principle of fabrication of a nanotrench device thanks to a tilted evaporation. b. Size of the nanotrench spacing as a function of the evaporation angle and thickness of the first electrode. Size below 20nm generally led to shorted devices due to the electrode roughness, while device above 100nm present a limited interest. c. SEM image of pristine nanotrench. d. SEM image of CdSe/CdS NPL functionalized nanotrench.*

As we used the same material as before (film of CdSe/CdS NPL) and switched from µm spaced electrodes to the nanotrench, we observed a dramatic effect on the photoresponse, which was boosted by more than 7 orders of magnitude (see Figure 24a). Only a factor 200 results from the increase of gain due to the shorter transit time. There are consequently other mechanisms responsible for this enhancement of the photoresponse. We have a stronger charge dissociation and the large applied electric field (500 kV.cm$^{-1}$) is now able to overcome the exciton binding energy. In other words, the bias energy drop per particle overcomes the binding energy. Secondly, with the nanoscale device the transport mechanism switches from hopping to a single tunnel event. This means that we removed a thousand tunnel events, and also that the carriers no longer experience the trap states. The photocharge lifetime is now limited by bimolecular recombination, as evidenced by the dependence of the photoresponse with the photon flux (see Figure 24a). The device can achieve a fast operation with a cut-off frequency measured at around 3 kHz (see Figure 24b-c).

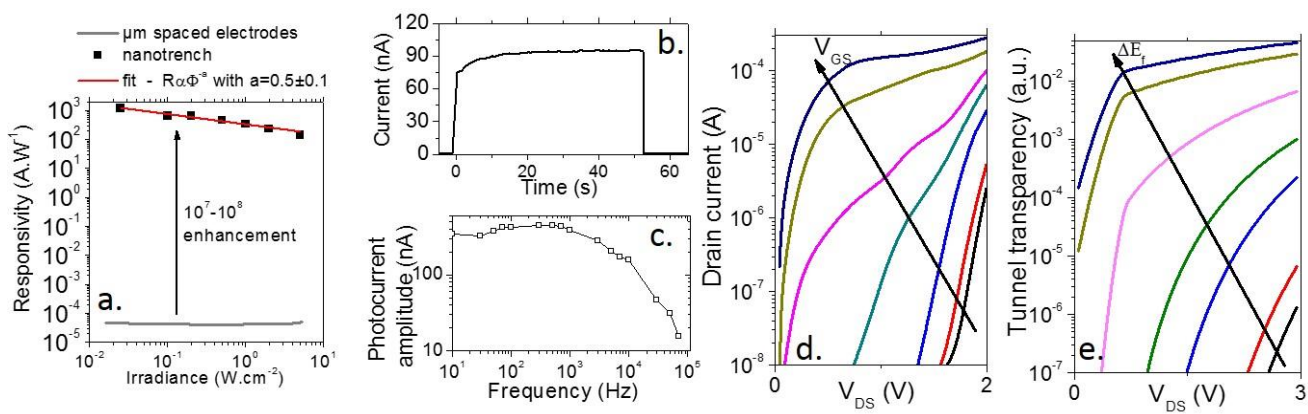

*Figure 24 a. Photoresponse as a function of the photon irradiance (λ=405nm) for a film of CdSe/CdS NPL deposited on 10µm spaced electrdes or on a nanotrench. b current in a CdSe/CdS NPL film deposited on a nanotrench under a pulse of light. c. Frequency dependence of the photocurrent for a CdSe/CdS NPL film deposited on a nanotrench. The cut-off frequency is around 3 kHz. d IV curves under different gate biases for a CdSe/CdS NPL film deposited on a nanotrench. e. Tunnel transparency relative to the injected contact as a function of the drain bias for different gate biases for a CdSe/CdS NPL film deposited on a nanotrench.*

We then coupled the CdSe/CdS film on the nanotrench with an ion-gel gate and observe a strong field effect, with on/off ratio of 10$^4$, mostly limited by the leakage in the ion gel (see Figure 24d). To clarify our understanding of transport in the nanotrench, we then compared these measurements with the tunnel transparency resulting from the Schottky barrier at the metal-NPL interface. The model includes the band alignment in the presence of confinement of the NPL with the metallic contact and a WKB approximation for the evaluation of the tunnel transparency. We obtained a fairly good qualitative agreement (see Figure 24e).

### 3.2.4. Metal-semiconductor hybride structure

During the internship of Loic Guillemot, we investigated the effect on transport of a gold functionalization on the CdSe NPL. This work was initially motivated by a report from the Murray group which introduced nanocrystals as dopant for a nanocrystal film of another kind[158,159]. There was also news from the Talapin group reporting charge transfer between gold and PbS[160], in an Au core /PbS shell heterostructure. Thus our initial goal was to find a path to change the carrier density within the NPL thanks to the addition of gold at a time when doping of the NPL was not possible. The growth of the gold tip on the CdSe NPL is actually inspired from the growth of the gold tip on CdSe rods[161,162]. Typically, a gold salt is reduced by amine and forms tips mostly located on the NPL corner (see Figure 25a-b). Thanks to a collaboration with Benoit Mahler, we then further expanded the gold functionalization. By using gold reduction under temperature and light illumination we demonstrated control of the site and size of the gold tips.

Unsurprisingly as we add the gold tip, the conductance of the metal semiconductor hybrid system rises (see Figure 25c-d). This phenomenon was also observed in Au functionalized gold tip[163,164], however the proposed explanations remained insufficient. Several mechanisms can be proposed. Gold might act as a conductive ligand. Indeed, we observe that in the presence of gold tips the CdSe NPL formed a network and the gold tips can bridge several NPL (see Figure 25a). However, this hypothesis can be ruled out since we still need a ligand exchange step to make the film conductive. A second mechanism may result from the formation of an $Au_2Se$ phase. Instead of forming two distinct phases of Au and CdSe, we may obtain a mixed material due to cation exchange[114]. Since $Au_2Se$ is a narrow band gap material, the thermally activated carrier density may increase, and this may result in the observed rise of conductance. To test this hypothesis, Francois Rochet from LPCMR, conducted XPS measurements. In the case of Au, Cd and Se only one oxidation state has been identified. And no $Au^+$ signature associated with an $Au_2Se$ phase was observed. Our third hypothesis, which we initially strongly believed was the good one, was an electron transfer from the gold to the CdSe NPL. However, neither the Cd core level shift from XPS, the change of Fermi level from UPS (made by A. Ouerghi at Tempo on synchrotron Soleil) nor the spatially resolved KPFM measurement confirmed this hypothesis (see Figure 25e-f). Indeed, this demonstrates the opposite effect, the Fermi level shift toward the valence band which suggest a hole filling[165] of CdSe. Since CdSe is only an n-type material[19,25], this actually means that CdSe is not involved in transport in the hybrid system. It thus brings us to the conclusion that only the gold is active on transport and that the rise of conductance results from a percolation process between gold islands.

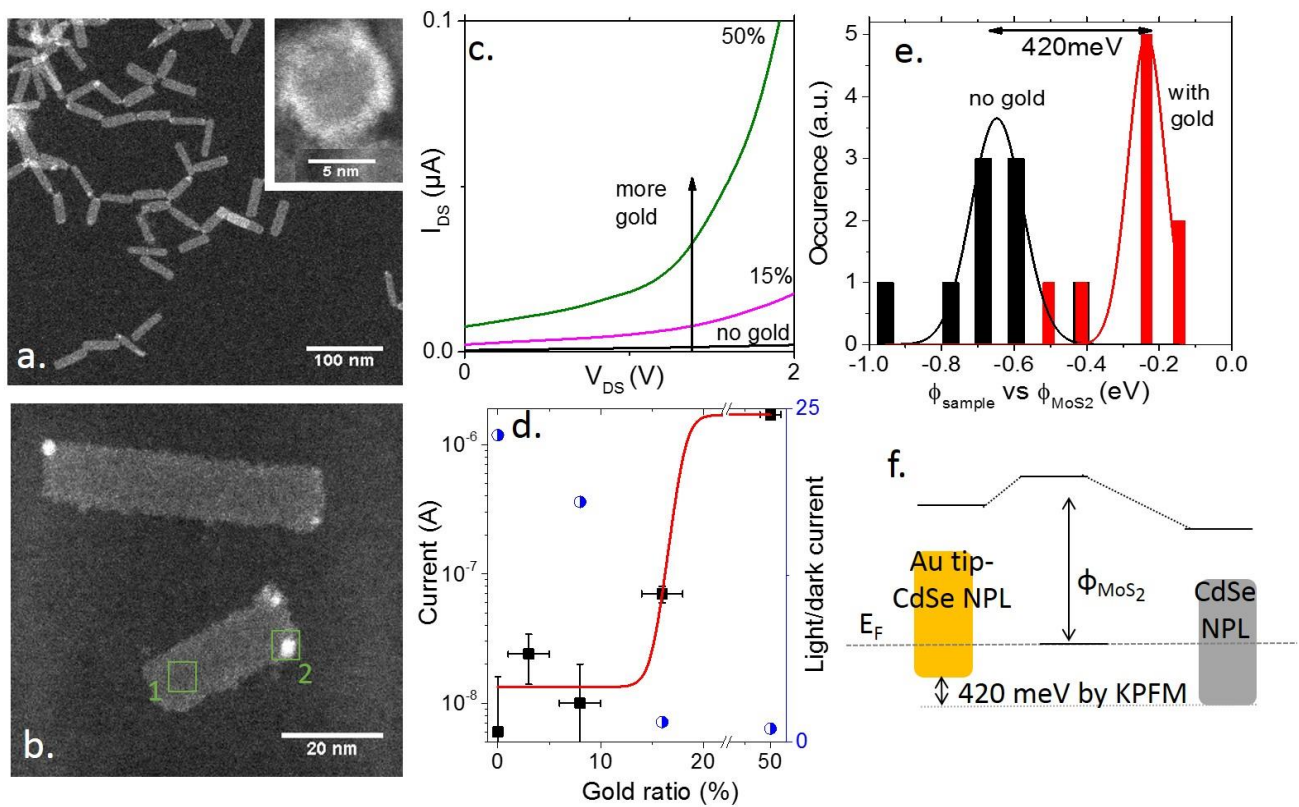

*Figure 25 a and b are TEM images of CdSe NPL functionalized by gold tips on the edge. c. IV curve relative to CdSe NPL functionalized with various amount of gold tips. d. Change of current and photocurrent as a function of the gold atomic ratio. e. Histogram of work function relative to the $MoS_2$ substrate work function for pristine CdSe NPL and gold tipped CdSe NPL measured by KPFM. f. Relative band structure of pristine CdSe NPL and gold tipped CdSe NPL.*

### 3.3. Conclusion

The use of anisotropic CQD for optoelectronics has indeed revealed that the shape and dimentionality of the object are key factors for photodetection. So far, our hope to significantly increase the mobility by reducing the number of hopping steps has not been fulfilled. However, we have identified that the shape strongly influences the dielectric screening and leads to a boost of the exciton binding energy, which has to be specially managed. My research now shifts toward the development and use of narrower band gap NPL. This work is conducted in collaboration with Sandrine Ithurria and is part of the PhD of Clément Livache.

# 4. VAN DER WAALS HETEROSTRUCTURES

Students involved in this work: Adrien Robin, Wasim Mir.

Publications relative to this work : 27, 29, 34,36.

### 4.1. Graphene nanoparticles Hybrids

This research activity had already started by the time we were investigating transport in NPL arrays. As I already indicated in the previous section, transport is a true challenge for colloidal nanocrystals. It was proposed in 2012 that one might use CQD as a graphene light sensitizer.[168] The basic idea is to take advantage of the large mobility of the graphene and the high absorption of the CQD. Absorption occurs within the semiconductor CQD, and thanks to a selective charge transfer a current can flow in the graphene. This idea is actually very similar to what can be done in dye-sensitized solar cells[166] and phototransistors[167].

#### *4.1.1. Motivation*

Since its (re)discovery in 2004, graphene has generated an impressive amount of research. This effort is motivated by the unique physical properties of graphene. Here, it is the electronic properties which raise our interest. Graphene is a semimetal with linear symmetric dispersion. As a result, electrons and holes present high carrier mobility. In the case of epitaxial graphene, values of 2000-4000 $cm^2V^{-1}s^{-1}$ are typical.

The coupling of the graphene layer with CQD really started with the work of Konstantatos *et al,*[168] followed by several other groups,[169] who demonstrated impressive photodetection performance in graphene-PbS CQD hybrid heterostructures. They reported extremely large photodetection gain and photoresponse up to $10^7$ $A.W^{-1}$. This work has paved the way for unconventional van der Waals heterostructures, where 2D materials are mixed with other low-dimensionality materials[170]. We revisited this approach during Adrien Robin's PhD while using CdSe NPL as a light sensitizer for graphene.

Thanks to a collaboration with Abdelkarim Ouerghi at C2N, we were provided with epitaxial graphene. This graphene combines two main practical advantages. It is immediately deposited on an insulating substrate and no transfer step is necessary. Secondly, the substrate is transparent so we can perform a back side illumination, which allows the building of a top gate. All operation of this hybrid device relies on the selective charge transfer between the dye and the transport layer. Under illumination, electron-hole pairs get generated in the semiconductor. If only energy transfer was occurring, the electron hole pair would quickly recombine in the graphene and only lead to heat. On the other hand, if one charge is preferentially transferred, the graphene carrier density is modulated and we observe its signature through a change of conductance. Ultimately, photodetection in this hybrid heterostructure is driven by the charge transfer. During the PhD of Adrien Robin we chose to investigate it thanks to a phototransistor configuration.

#### *4.1.2. On demand charge transfer*

The graphene layer on SiC substrate is first patterned via three steps of lithography. On the top of it we deposit CdSe NPL and finally electrolyte is deposited on the top of the whole heterostructure. A side gate is used to control the carrier density. We first conducted the transistor measurements without NPL. The measured transfer curve is shown in Figure 26a. This conductance curve is typical for graphene and presents a minimum which corresponds to the bias where the Fermi level crosses the Dirac point. As is, the graphene is already n-type which results from an electron transfer from the SiC substrate to the graphene.[171] Once NPL are added, the Dirac point shifts to positive gate biases. This shift, of almost 1 V, corresponds to an electron transfer from the graphene to the NPL of ≈$10^{13}$ $cm^{-2}$ carriers, assuming a gate capacitance of 1-2 µF.$cm^{-2}$ which is typical for electrolytically gated graphene.[172,173] Under illumination, the Dirac point shifts reversibly back toward negative bias. Under illumination, electron-hole pairs get generated in the n-type NPL. The electrons are transferred to the graphene, while the holes remain trapped within the semiconductor. Thanks to the large mobility of the graphene, the electron can recirculate several times during the lifetime of the hole. The latter is likely limited by its recombination with environment, as indicated by the sensitivity of the system to air and moisture. This gain mechanism is actually the same as the one occurring in the pure film of NPL.[19]

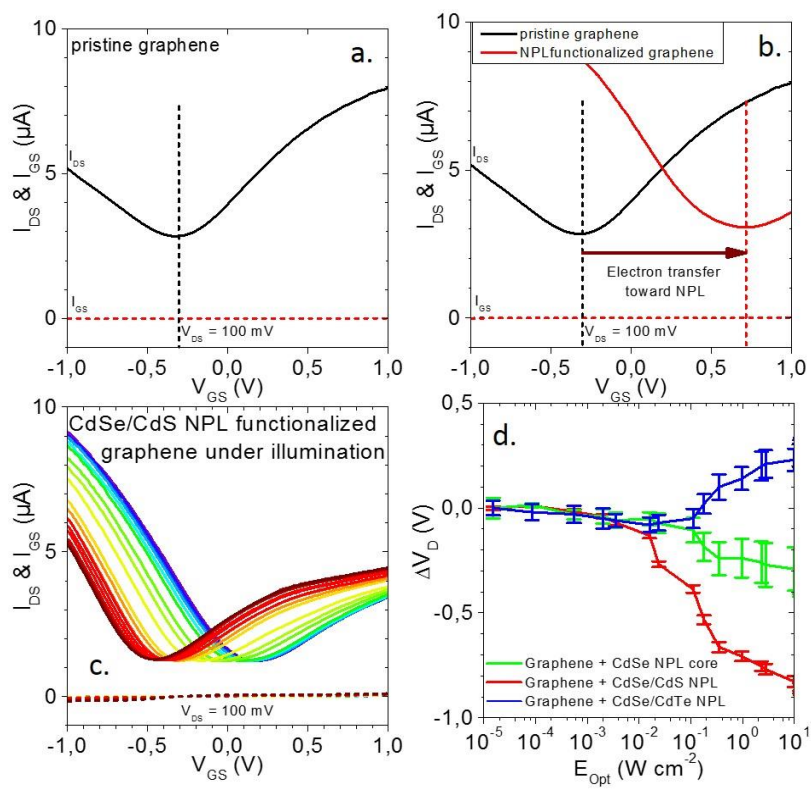

*Figure 26 a Transfer curve (drain current vs gate bias) for a pristine graphene channel. b Transfer curve for a CdSe/CdS NPL functionalized graphene channel. c. Transfer curve for a CdSe/CdS NPL functionalized graphene channel under different level of illumination. d. change of the minimum of conductance gate bias as a function of the light irradiance (λ=405 nm) for CdSe NPL functionalized, for a CdSe/CdS core shell NPL functionalized and for a CdSe/CdTe core crown NPL functionalized graphene channel.*

The charge transfer between the NPL and the graphene has to be carrier selective to prevent energy transfer. Actually, both charge and energy transfer occur but generally only one of them is probed, either charge transfer by transport measurement or energy transfer for FRET measurement. This selectivity of the charge transfer relies on two key parameters, which are the band offset of the semiconductor with graphene and the binding energy. The large binding energy of the NPL is not so favorable to an efficient charge dissociation. This explains why at the end the NPL-based hybrid presents lower performance than the PbS-based hybrid. Indeed, in PbS, the carriers almost behave as free charges due to the large dielectric constant of PbS which makes the binding energy very weak (few meV). To boost the charge transfer, we then need to find a way to decrease the binding energy. This can be done by reducing the overlap between the electron and the holes. In the III-V semiconductor heterostructure, such a strategy has already been proposed to induce dipole.[174] We chose to use colloidal hetrostructures as a way to reduce the coulombic interaction between electron and hole. Two types of heterostructures have been tested, CdSe/CdS core-shell and CdSe/CdTe core-crown heterostructures. For core-shell, the CdSe NPL is encapsulated in a CdS shell and it is mostly the confined direction (i.e., the thickness) which is affected. The electron gets localized all over CdSe and CdS, while the hole remains confined in CdSe. For CdSe/CdTe core-crown, the thickness of the core remains unaffected and the growth of the CdTe occurs only in the plane. In this case, the electron stays in CdSe, while the hole gets localized in the CdTe crown. The overlap between electron and hole gets reduced and the rise of the PL lifetime is a signature of the reduced overlap. As we switch from core to core-shell, we obtain an increase of the charge transfer by a factor 3 (see Figure 26d). On the other hand, as type II (CdSe/CdTe) band alignment is used, we observe that holes are now injected into the graphene. We thus have demonstrated a magnitude- and sign-tunable charge transfer from semiconductor to graphene. We now hope that the proposed strategy can be further used for the design of van der Waals heterostructures with a high level of control of the inter layer charge transfer.

### 4.1.3. Noise in the hybrid structures

In this hybrid heterostructure the photoresponse can be large thanks to gain, but several issues have been swept under the rug. In particular, the photoresponse almost scales like the inverse of the photon flux. In other words, good performance is limited to very low photon flux. In our hybrid the generated photocurrent remains unchanged over six orders of magnitude of photon flux. Once the electron-hole pair is generated the hole gets trapped and will then behave as a recombination center for the next photogenerated electron, and that leads to a huge non-linearity of the photoresponse.

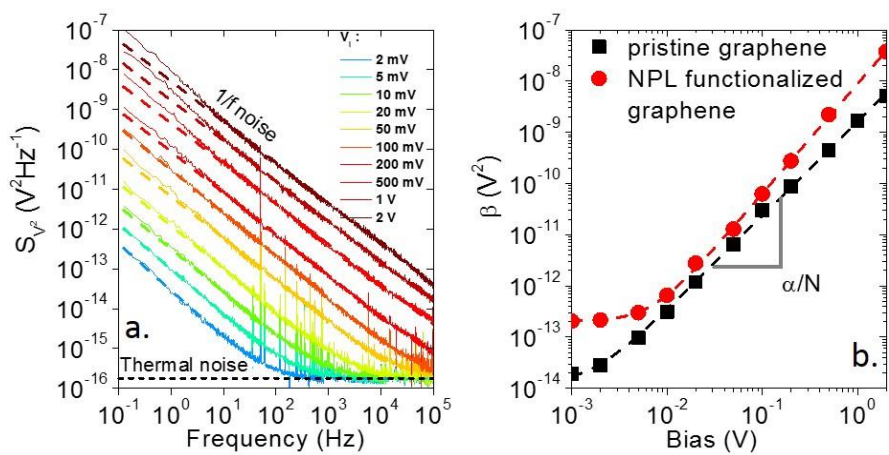

*Figure 27 a. Noise voltage density as a function of signal frequency for a graphene channel under different applied biases. b. Relative magnitude of the 1/f noise contribution of the noise voltage density as a function of the applied bias for a pristine graphene channel and for a CdSe/CdS NPL functionalized graphene channel.*

A second key issue that needs to be addressed in this hybrid device is whether or not the presence of the nanoparticles brings more noise. To measure the noise in this device, Adrien Robin has built a setup where four devices are connected in a Wheatstone bridge configuration.[27] This allows reducing most of the DC part of the signal, which can then be amplified and acquired on a spectrum analyzer. The voltage spectral density presents two clear components—a white noise, which we attribute to the thermal noise, and a huge contribution due to *1/f* noise[175] (see Figure 27a). Actually, in most published papers on this subject, this contribution, which cannot be estimated *a priori* by an analytical expression, is simply neglected. As a result, the detectivity values provided in many papers are largely overestimated. The magnitude of the 1/f noise is then compared for pristine graphene and NPL functionalized graphene. We measure a rise of the noise by a factor 3 (see Figure 27b). However, this difference can be fully attributed to the difference of carrier density within the graphene, according to the *1/N* dependence given by Hooge's law. The conclusion here is that functionalizing the graphene tunes the noise according to the change in graphene carrier density.

Performance-wise these devices are not fully satisfying. The gain of performance resulting from the higher mobility is not balanced by the rise in the carrier density, due to the lack of band gap in graphene. This explains why most recent projects have switched from graphene to TMDC material such as $MoS_2$.[176] Another limitation of this system, which is easy to point *a posteriori,* is the fact that the photocurrent modulation in graphene cannot be more that what can be obtained on the transfer curve. As shown in Figure 26, we have obtained a modulation by a factor 4. A photomodulation of a factor of 4 while using a wide band gap semiconductor is very unimpressive, since we already demonstrated modulation by 4-5 orders of magnitude in pure film of NPL. On the other hand, revisiting this strategy in the infrared might be worthwhile. In the MWIR or LWIR, having photomodulation of only a few percent is common. The coupling of graphene with mid-IR CQD looks far more promising, since in this case we can expect some improvements.

### 4.2. 2D Heterostructure: demonstration of a pn junction in graphene $MoS_2$

After this first work on graphene, we pushed our collaboration with A. Ouerghi to the next step and investigated transport in more conventional van der Waals heterostructures. The device is a stack of p-graphene with n-type $MoS_2$, as shown in Figure 28a-b. Briefly, a $MoS_2$ synthetically grown by CVD is transferred on a hydrogenated epitaxial graphene. Contacts are then deposited after an e-beam lithography step. All the steps of this process were performed by the C2N team, while the transport and phototransport experiments were conducted on our transport experimental setup.

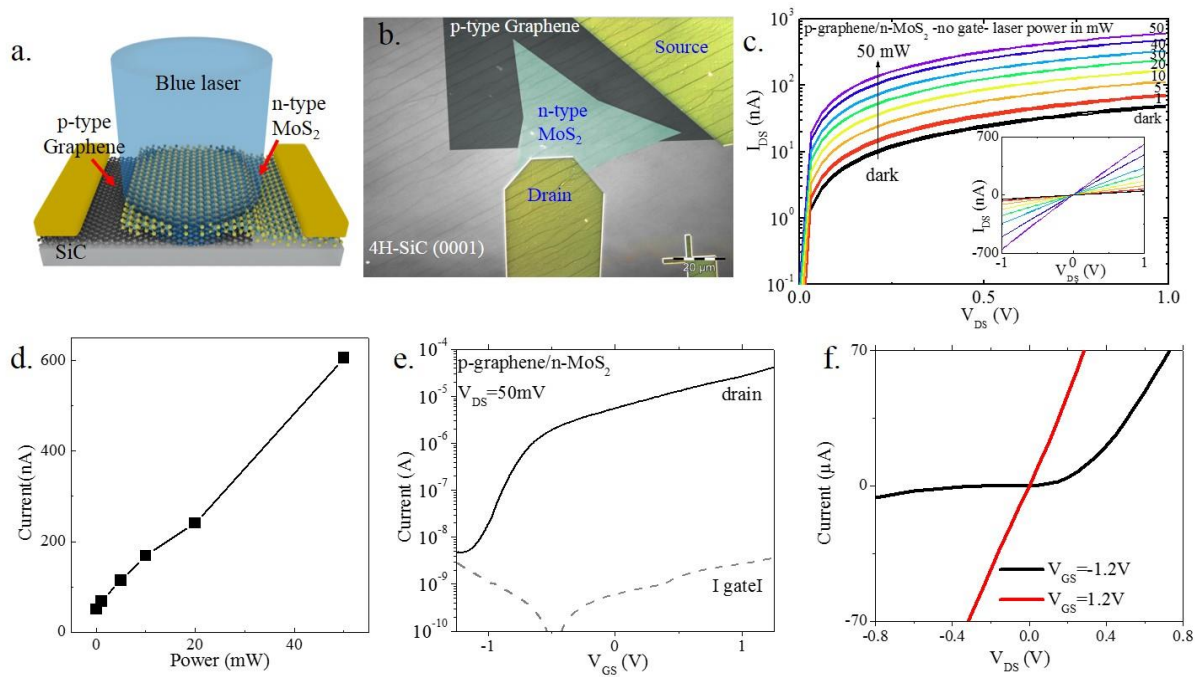

*Figure 28 a and b are repectively a scheme and false color microscopy image of a graphene/$MoS_2$ vertical jucntion. c. IV curves under different light irradiance for the graphene/$MoS_2$ vertical jucntion. The inset is the same set of curves in linear scale. The power dependence of the photocurrent is quite linear as shown in part d. e. Transfer curve of the graphene/$MoS_2$ vertical jucntion. f. IV curve of the graphene/$MoS_2$ vertical jucntion under hole ($V_{GS}<0$ V) and electron injection ($V_{GS}>0$ V).*

The device is photoresponsive and the current rises linearly with the photon flux (see Figure 28c-d). We then added a top ion gel gating to investigate the impact of the Fermi level tuning. The transfer curve shows that the junction behaves as an effective n-type material (see Figure 28e). The on-off ratio reached almost four orders of magnitude. Even more interesting is the change of shape of the IV curve under gate control. Under electron injection ($V_{GS}$>0 V) the IV curve is linear. In fact, the graphene is barely p-type, and under electron injection the Fermi level is moved above the Dirac point. We obtain a n-n junction with an ohmic behavior. On the other hand, if holes are injected, the p-character of the graphene is reinforced, while the inherent n-doping of the $MoS_2$ is high enough to preserve its n-type nature. We thus form a pn junction which is responsible for the rectifying behavior of the I-V curve. In this study we also demonstrated that the sign and magnitude of the photoresponse are gate tunable.

These mixed-dimensionality van der Waals heterostructures are models for the investigation of charge transfer processes. So far our probe (i.e., transport under gate control) is well-suited to probe the static charge transfer. My goal will be now to gain insight on the dynamic of the coupling between layers.

# 5. PERSPECTIVES

This section has been removed.

# 6. REFERENCES

## Publications in peer review journal

## Conference proceedings

## Patents and patents applications

## INVITED TALKS

## SEMINARS in LABORATORY

## Other publications